\documentclass[fleqn,usenatbib]{mnras}
\usepackage{newtxtext,newtxmath}
\usepackage[T1]{fontenc}

\DeclareRobustCommand{\VAN}[3]{#2}
\let\VANthebibliography\thebibliography
\def\thebibliography{\DeclareRobustCommand{\VAN}[3]{##3}\VANthebibliography}

\usepackage{CJKutf8}
\usepackage{booktabs}
\usepackage{xcolor, soul}

\usepackage{amsmath, amssymb, bm} 
\usepackage{graphicx}
\graphicspath{{figures/}}

\title[Cosmology with Scattering Transform]{A new approach to observational cosmology using the scattering transform}

\author[Cheng, Ting, M\'enard, \& Bruna]{
Sihao Cheng (程思浩),$^{1}$\thanks{E-mail: s.cheng@jhu.edu}
Yuan-Sen Ting (丁源森),$^{2,3,4,5}$
Brice M\'enard,$^{1}$
and Joan Bruna$^{6,7,3}$
\\
$^{1}$Department of Physics and Astronomy, The Johns Hopkins University, 
Baltimore, MD 21218, USA\\
$^{2}$Institute for Advanced Study, Princeton, NJ 08540, USA\\
$^{3}$Department of Astrophysical Sciences, Princeton University, Princeton, NJ 08544, USA\\
$^{4}$Observatories of the Carnegie Institution of Washington, 
Pasadena, CA 91101, USA\\
$^{5}$Research School of Astronomy \& Astrophysics, Australian National University, 
Weston, ACT 2611, Australia\\
$^{6}$Courant Institute of Mathematical Sciences, New York University, New York, NY 10012, USA\\
$^{7}$Center for Data Science, New York University, New York, NY 10011, USA
}

\date{Accepted 2020 October 07. Received 2020 September 24; in original form 2020 July 15}
\pubyear{2020}

\begin{document}
\label{firstpage}
\pagerange{\pageref{firstpage}--\pageref{lastpage}}
\begin{CJK}{UTF8}{gkai} 
\maketitle
\end{CJK}

\begin{abstract}
Parameter estimation with non-Gaussian stochastic fields is a common challenge in astrophysics and cosmology. In this paper, we advocate performing this task using the \emph{scattering transform}, a statistical tool 
sharing ideas with convolutional neural networks (CNNs) but requiring no training nor tuning. It generates a compact set of coefficients, which can be used as robust summary statistics for non-Gaussian information. It is especially suited for fields presenting localized structures and hierarchical clustering, such as the cosmological density field.

To demonstrate its power, we apply this estimator to a cosmological parameter inference problem in the context of weak lensing. On simulated convergence maps with realistic noise, the scattering transform outperforms classic estimators and is on a par with state-of-the-art CNN. It retains advantages of traditional statistical descriptors, has provable stability properties, allows to check for systematics, and importantly, the scattering coefficients are interpretable. It is a powerful and attractive estimator for observational cosmology and the study of physical fields in general.\\
\end{abstract}


\section{Introduction}
\label{sec:intro}

Non-Gaussian fields are ubiquitous in astrophysics. Analysing them is challenging, as the dimensionality of their description can be arbitrarily high. In addition, there is usually little guidance on which statistical estimator will be most appropriate for parameter inference. In this paper, we advocate using a novel approach, called the scattering transform \citep{Mallat_2012}, for the analysis of such fields and, in particular, the matter distribution in the Universe, a highly studied non-gaussian field.

In many areas of astrophysics and, in particular, in cosmology, extracting non-Gaussian information has been attempted through $N$-point correlation functions \citep[e.g.,][for weak lensing applications]{Bernardeau2002,Takada_2003, Semboloni_2011, Fu_2014} and polyspectra, their Fourier equivalents \citep[e.g.,][]{Sefusatti_2006}. Correlation functions are convenient for theoretical predictions and for measuring weak departures from Gaussianity. However, being high powers of the input field, these statistics suffer from an increasing variance and are not robust to outliers in real data, making them gradually less informative \citep{Welling_2005}. If the distribution of field intensity has a long tail, the amount of information accessible to $N$-point functions will quickly decrease \citep{Carron_2011}. In addition, the number of configurations to consider for $N$-point functions explodes with the number of points used. As a result, information is highly diluted among coefficients, and it becomes a challenge to efficiently extract information with $N$-point functions. Other methods, including performing a non-linear transformation before calculating correlation functions \citep{Neyrinck_2011, Simpson_2011, Carron_2013, Giblin_2018}, using topological properties such as Minkowski functionals \citep{Mecke_1994, Hikage_2003, Shirasaki_2014, Kratochvil_2012}, and using biasing properties such as counts of clusters, peaks, and voids \citep{Jain_2000, Marian_2009, Kratochvil_2010, LPH15, LPL15, Pisani_2019}, have also been considered. However, in the cosmological context, these excursions into non-Gaussian signal analyses have had limited impact in improving existing constraints on cosmological parameters so far.

Recently, convolutional neural networks \citep[CNNs, e.g.,][]{lecun-98} have claimed supremacy in a wide variety of applications aimed at extracting information from complex data. They have also shown promises to efficiently retrieve cosmological information well beyond second-order statistics \citep[e.g.,][]{Gupta_2018, Ribli_2019a, Ribli_2019}. While the potential of this method is enormous, it also comes with a number of issues. To precisely and robustly estimate cosmological parameters, CNNs require a large training set. In addition, when applied to real data, systematic errors not included in the training process of CNN can hardly get checked and controlled, whereas for traditional statistics, a simple $\chi^2$ test can do so. As such, the use of CNNs in real data comes with limitations regarding interpretability and validity.

In this paper, we advocate using a different approach called the \textit{scattering transform} to efficiently and robustly extract statistical information from non-Gaussian fields\footnote{This work was done simultaneously and independently of that presented in \cite{Allys_2020}, where the authors apply a different but related technique, the wavelet phase harmonic, to slices of the matter density field.}. The operations and structure of the scattering transform has close similarities with those built in CNNs, but the scattering transform does not require any training, and like traditional statistics, it generates coefficients with proved properties. It can therefore hopefully overcome the aforementioned limitations encountered with CNNs. In section~\ref{sec:ST}, we introduce the scattering transform, present intuitive understanding of its coefficients, and visualise its key properties. In section~\ref{sec:WL_application} and \ref{sec:results}, we demonstrate the power of the scattering transform to infer cosmological parameters ($\Omega_\text{m}$ and $\sigma_8$)
in the context of weak lensing using simulated convergence maps. As we will show, it outperforms the power spectrum and peak counts, and is on par with the state-of-the-art CNN. Finally, we comment on the attractive properties of the scattering transform in Section~\ref{sec:discussion} and conclude in Section~\ref{sec:conclusion}.

\section{The scattering transform}
\label{sec:ST}

In this section, we present the scattering transform and intuitive interpretations of its coefficients. The scattering transform generates a compact set of coefficients that captures substantial non-Gaussian information beyond the power spectrum. In contrast to $N$-point functions, the scattering coefficients are all proportional to the input data, and do not suffer from the increasing variance issue. Thus, the scattering coefficients, which form a representation of the input field, can be used to extract non-Gaussian information efficiently and robustly. This is particularly attractive from a data analysis point of view.

\subsection{Motivation}
\label{sec:st_motivation}

The scattering transform was originally proposed by \citet{Mallat_2012} as a tool for signal processing to extract information from high-dimensional data. 
In contrast to neural networks, it comes with attractive provable properties including translational invariance, non-expanding variance, and Lipschitz continuous to spatial deformation \citep{Mallat_2012}. Interestingly, the scattering transform has also provided key insights into deciphering the remarkable behaviour and performance of CNNs \citep{Bruna_2013}. A perhaps counter-intuitive feature of CNNs is that the convolution, though restricting the flexibility of the neural network, dramatically boosts its performance on many types of data. In addition, a successful CNN architecture can often be re-purposed for very different tasks. These facts suggest that a certain mathematical structure enables efficient information extraction from a wide range of complex data. Understanding this structure may dramatically simplify the costly training process required when using neural networks.

The scattering transform has been successfully used in many areas, including audio signal processing \citep{AndenMallat_2011, AndenMallat_2014}, image classification \citep{Bruna_2013}, texture classification \citep{Sifre_2013}, material science \citep{Hirn_2017, Eickenberg_2018, Sinz_2020}, multifractal analysis in turbulence and finance \citep{Bruna_2015}, and graph-structured data \citep{gama2018diffusion}. Several of these examples reached state-of-the-art performance compared to the CNNs in use at the time. In astrophysics, a pioneer application has been performed by \citet{Allys_2019} to analyse the interstellar medium. 

The scattering transform can be used with two possible goals in mind: representing a specific realisation of a field (with a classification goal) or characterising the global statistical properties of a field. The narratives in these two regimes are slightly different \citep{Mallat_2012, Bruna_2013}. We will focus on the latter one, which is relevant to cosmological applications.

\begin{figure*}
    \centering
    \includegraphics[width=0.65\textwidth]{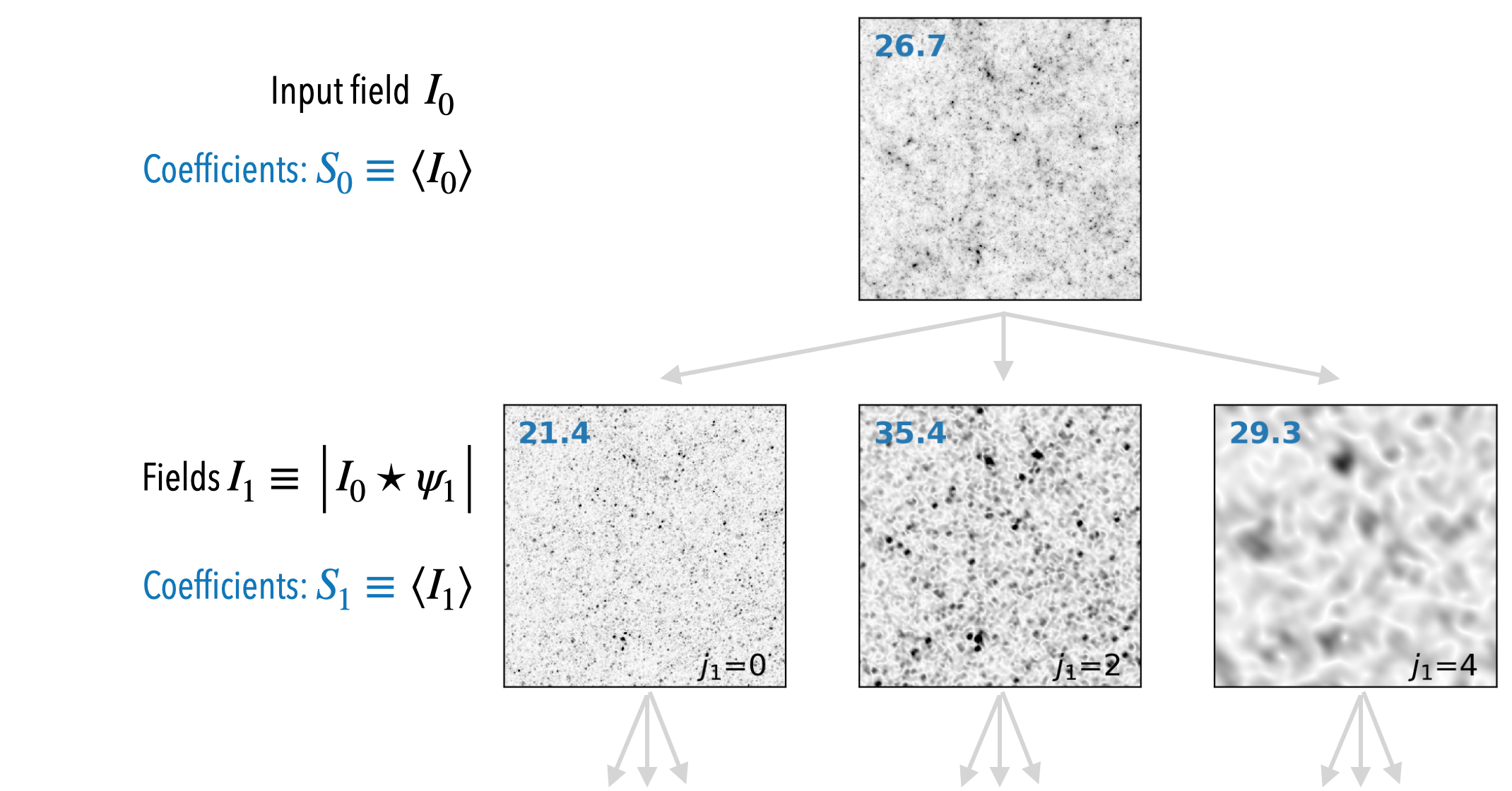}
    \includegraphics[width=0.65\textwidth]{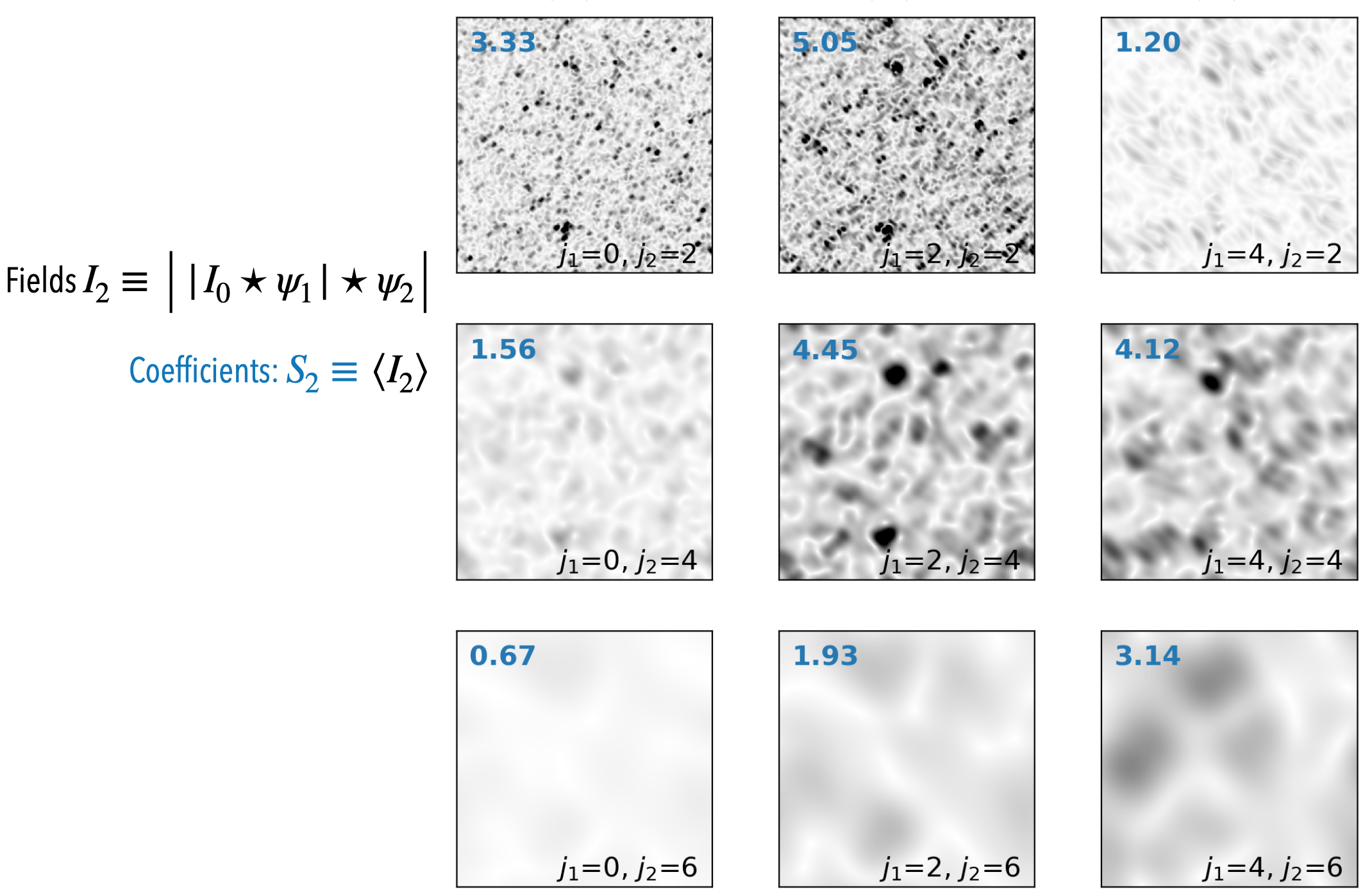}
    \caption{Illustration of the scattering transform on a weak lensing map. The azimuthal resolution is set to be $L$ = 4. For clarity, we only show results using wavelets with orientation indices $l_1$ = 1 and $l_2$ = 1, and several selected scale indices $j_1$ and $j_2$. In the top left corner of each panel, we show the mean value of that field. They are the scattering coefficients ($S_0$, $S_1$, $S_2$) of the input field. For a convenient display, the blue numbers are 10$^4$ times the coefficients derived from the lensing map. E.g., the $S_0$ coefficient of this lensing map is actually 0.00267. The color bar ranges for $I_0$, $I_1$, $I_2$ fields are adjusted separately for better visualization.}
    \label{fig:I}
\end{figure*}

\begin{figure}
    \centering
    \includegraphics[width=\columnwidth]{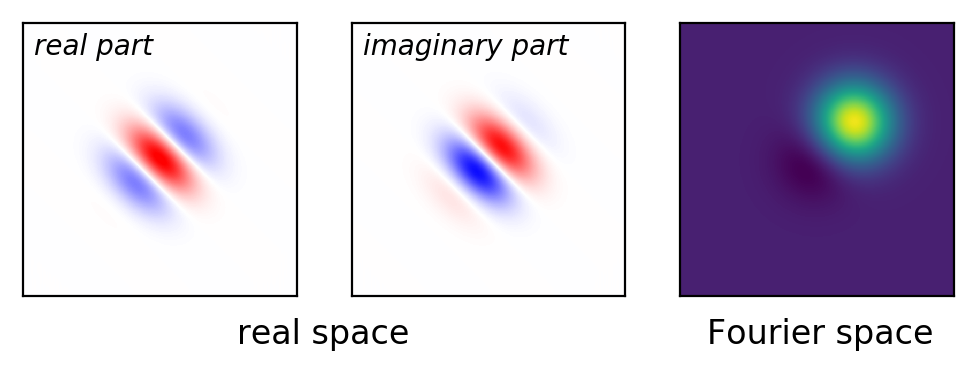}
    \includegraphics[width=\columnwidth]{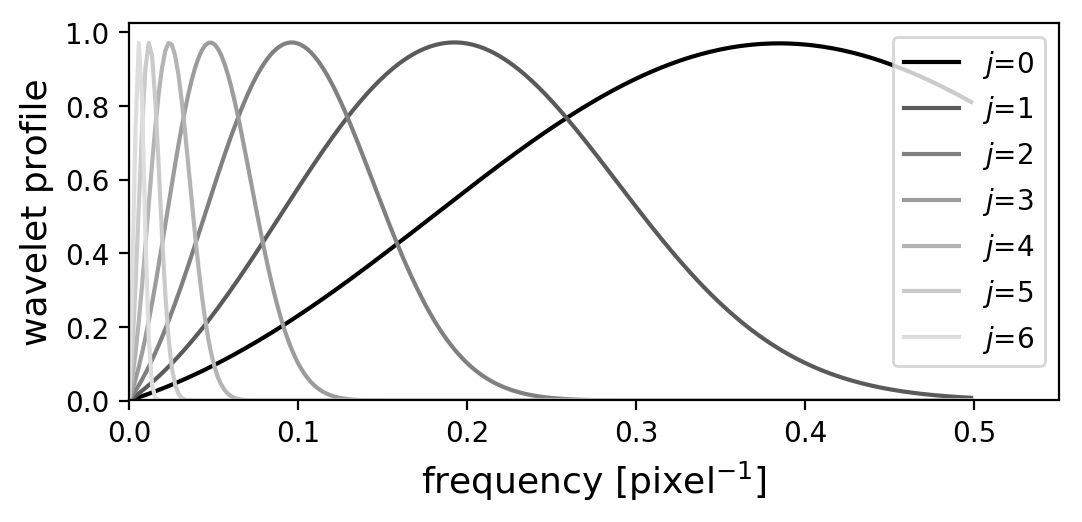}
    \caption{\textit{Upper panel}: 
    profile of a Morlet wavelet ($j$~=~6, $l$~=~0, image size 512$\times$512 pixels) in the real space and another one ($j$ = 1, $l$ = 1) in Fourier space. The centre of the Fourier space represents zero frequency.
    \textit{Lower panel}: radial frequency profiles of a family of wavelets. Dilating/contracting (by factor of 2) and rotating (by $\pi$/L) one wavelet give the whole family of wavelets used in the scattering transform. 
    }
    \label{fig:wavelets}
\end{figure}

\subsection{Formulation}
\label{sec:st_formulation}

Here, we present the formulation of the scattering transform in the context of characterizing random fields \citep{Mallat_2012}. We focus on the 2-dimensional case in this study, but it can be directly generalized to any other dimensionality. For clarity, we will attach the notation $(x,y)$ for the spatial dependence of a field only when it is first introduced.

To extract information from an input field, the scattering transform first generates a group of new fields by recursively applying two operations: a wavelet convolution and a modulus. Then, the expected values of these fields are defined as the scattering coefficients and used to characterize statistical properties of the original field (see Figure~\ref{fig:I} for an illustration). This hierarchical structure, the use of localized convolution kernels, and the use of non-expansive non-linear operator are all elements found in the architecture of CNNs.

Formally, given an input field $I_0(x,y)$, the scattering transform generates a set of 1st-order fields $I_1(x,y)$ by convolving it with a family of wavelets $\psi^{j,l}(x,y)$ and then taking the modulus:
\begin{align}
    I_1 &\equiv \left|I_0 \star \psi^{j_1,l_1}\right|\,,
\label{eq:I1}
\end{align}
where $I_1$ represents a group of fields labelled by the wavelet index $j_1$, $l_1$. Wavelets are localized oscillations and band-pass filters. Figure~\ref{fig:wavelets} shows the profiles of Morlet wavelets in real and Fourier space. Morlet wavelets are used in our study and described in Appendix \ref{app:Morlet}. In general, a family of wavelets covers the whole Fourier space. They all have the same shape but different sizes and orientations, labelled by $j$ and $l$ respectively. They can all be generated through dilating and rotating a prototype wavelet. In the scattering transform, the convention is to use a dilation factor of 2, such that for a pixelized field, the size of a wavelet $\psi^{j,l}$ in the real space is roughly $2^j$ pixels.

Having created the 1st-order fields, one can then iterate the same process to create 2nd-order fields $I_2(x,y)$:
\begin{align}
    I_2 &\equiv \left|I_1 \star \psi^{j_2,l_2}\right|\nonumber\\ 
    & = \left||I_0 \star \psi^{j_1,l_1}|\star\psi^{j_2,l_2}\right|\,,
\end{align}
where $I_2$ represents a group of fields labelled by the two sets of wavelet index $j_1$, $l_1$ and $j_2$, $l_2$. An illustration of these first two orders of scattering transform is shown in Figure~\ref{fig:I}. Higher-order scattering fields can be created with further iterations. We note that the iterations are not commutative, so maintaining the order of wavelets is important. In this paper, we will only show the scattering transform up to the 2nd order, because we find that in our particular dataset, little cosmological information is stored in the 3rd order. 

If the input field $I_0$ is homogeneous, then all the generated fields $I_n$ remain homogeneous. Therefore, the expected values of their intensity can be used as translation-invariant descriptors of the input field:
\begin{align}
    S_0 &\equiv \langle I_0 \rangle \label{eq:S0}\\
    S_1^{j_1,l_1} &\equiv \langle I_1^{j_1,l_1}~~~~~~~\rangle = \langle \left|I_0\star\psi^{j_1,l_1}\right| \rangle \label{eq:S1}\\
    S_2^{j_1,l_1,j_2,l_2} &\equiv \langle I_2^{j_1,l_1,j_2,l_2}\rangle = \langle  \left| |I_0\star\psi^{j_1,l_1}|\star\psi^{j_2,l_2}\right| \rangle\,. \label{eq:S2}
\end{align}
These expected values $S_n$ are called the $n$th-order scattering coefficients. Due to homogeneity, these expected scattering coefficients can be estimated by taking the spatial average of a single realization:
\begin{equation}
    \hat{S}_n=\langle I_n \rangle _{x,y}\,,
\end{equation}
where $\hat{S_n}$ is an unbiased estimator of $S_n$ and $\langle \cdot \rangle _{x,y}$ represents the spatial average of a field.\footnote{To follow the convention in cosmology, we use slightly different notations from \citet{Mallat_2012}: we use $S_n$ to represent the expected values, which characterize properties of a random field and which Mallat denotes as $\overline S_n$; we use $\hat{S_n}$ to represent $S_n$'s estimators calculated from spatial average, which Mallat directly denotes as $S_n$.}

The number of scattering coefficients $S_n$ is determined by the number of wavelets used. Setting $J$ different scales ($2^J$ cannot exceed the side length of the field) and $L$ different orientations results in $J\times L$ different wavelets used in total. If all combinations of wavelets are used, then the number of coefficients at the $n$th-order will be $J^n L^n$. 

\begin{figure*}
    \centering
    \includegraphics[width=0.62\textwidth]{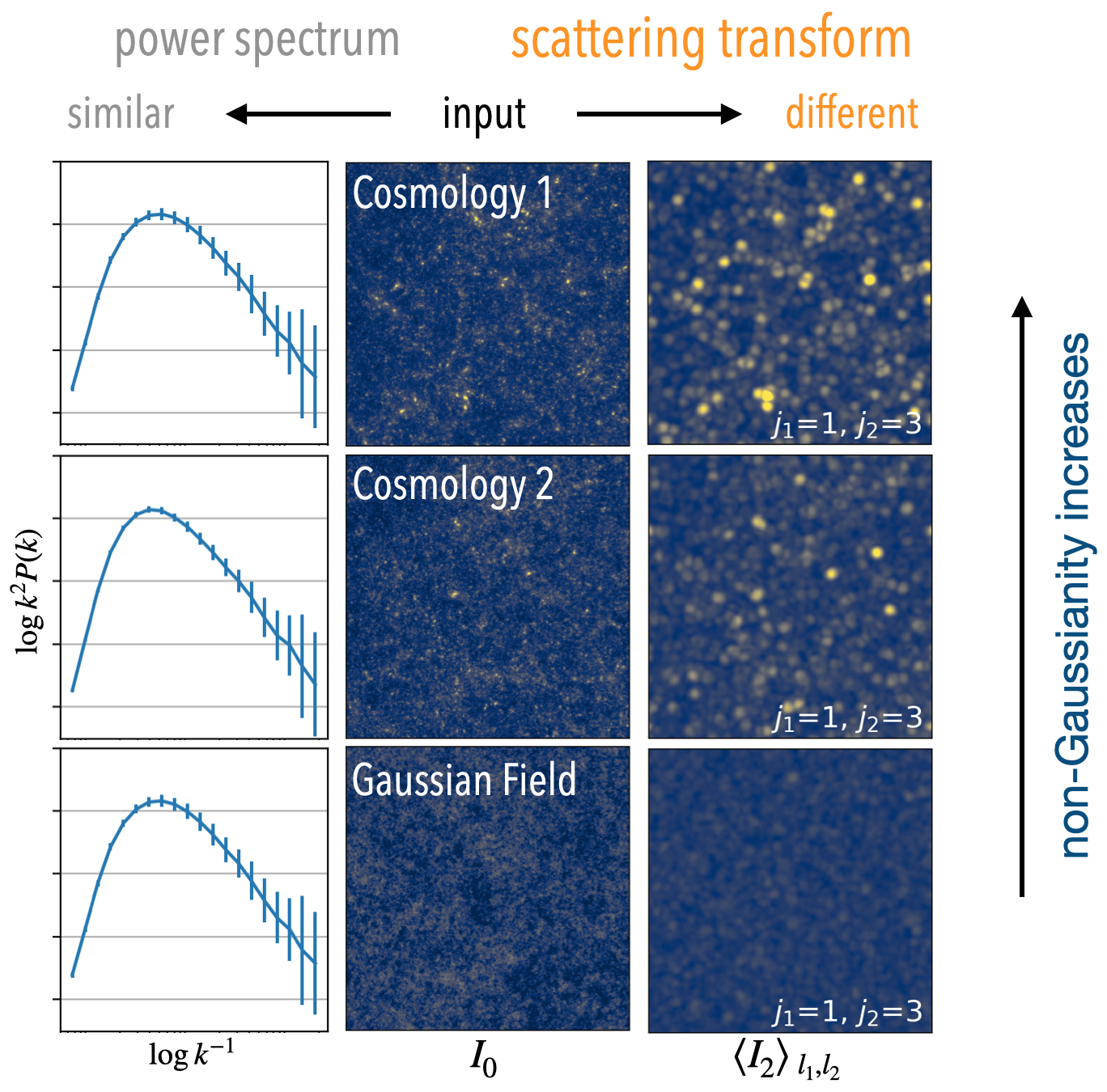}
    \caption{
    The scattering transform of three fields ($I_0$) with indistinguishable power spectra. 
    Row 1 shows a realization of convergence maps in cosmology $(\Omega_\text{m}, \sigma_8)$ = (0.292, 0.835), row 2 shows cosmology $(\Omega_\text{m}, \sigma_8)$ = (0.566, 0.520), row 3 is for a Gaussian random field with the same (2D) power spectrum as row 1. It can be seen by eye that the average intensity of the 2nd-order scattering fields (the last column), which corresponds to an $s_2$ coefficient and measures the clustering strength of structures highlighted by $I_1$, is significantly different from each other, while their power spectra (the first column) are indistinguishable.
    \label{fig:cos_diff}}
\end{figure*}

When considering an isotropic field, which is the case of interest in cosmology, the scattering coefficients $S_n$ can be further reduced. To construct isotropic statistics, we simply average over all orientation indices, which reduces the number of coefficients by an order of $L^n$ and creates a more compact and robust set of statistical descriptors. We thus define our reduced scattering coefficients as
\begin{align}
    s_0 & \equiv S_0\\
    s_1^{j_1} &\equiv \langle S^{j_1,l_1}_1\rangle _{l_1}\\
    s_{2}^{j_1,j_2} &\equiv {\langle S^{j_1,l_1,j_2,l_2}_2 \rangle _{l_1, l_2}}\,,
\end{align}
where $S_n$ represent the standard scattering coefficients, $s_n$ represent our reduced coefficients, and $\langle \rangle_l$ denotes an average over orientation indices. The reduced coefficients $s_n$ can also be understood as the expected value of some `reduced' fields $\langle I_n \rangle_{l_1,..,l_n}$, which are `stacks' of the $I_n$ with same scale indices $j$ but different orientation indices $l_1,..,l_n$. We show several examples of 2nd-order `reduced' fields in Figure~\ref{fig:cos_diff}, where information is condensed, so features look clearer than in Figure~\ref{fig:I}. Our reduction is similar to the first group of isotropic coefficients used by \citet{Allys_2019}. Up to the 2nd order, our reduced set includes $1+J+J^2$ coefficients. As a result, probing the full range of scales for an image with 512$\times$512 pixels ($J$ = 8) yields in total 73 reduced scattering coefficients. 

In general, performing azimuthal averages over both $l_1$ and $l_2$ leads to information loss. To preserve more isotropic information, one could keep $l_2-l_1$ as an index of the reduced 2nd-order scattering coefficients \citep{Bruna_2013, Allys_2019} or apply the `scattering strategy' again to rotation \citep{Sifre_2013}. In the weak lensing study presented below, however, we checked that this additional information does not improve the performance of our analysis, probably due to the lack of anisotropic structures in the weak lensing maps we use. So, we do not take it into account.

Having introduced the mathematical formulation of the scattering transform, we will present in the next section some intuitive understanding of its key operations.

\subsection{The role of wavelet convolution and modulus}
\label{sec:convolution and modulus}

The core operation $I\rightarrow|\I\star\psi^{j,l}|$ employed by the scattering transform comprises two steps: a convolution by a complex-valued wavelet and a modulus operation. In short, the wavelet convolution selects scales, and the modulus converts fluctuations into their local strength.

Let us discuss the wavelet convolution first. As a wavelet is a band-pass filter, the wavelet convolution selects Fourier modes around a central frequency and coarsely separates information of different scales (see Figure~\ref{fig:wavelets}). Due to the locality of wavelets in real space, which is related to their logarithmic spacing and widths in Fourier space, the scattering coefficients are Lipschitz continuous to deformation, meaning that similar fields differing by a small deformation (including a small dilation) are also similar in the representation formed by scattering coefficients \citep{Mallat_2012}, and therefore the scattering characterization is a stable one. Fourier coefficients (without binning), in contrast, are not stable to deformation at high frequencies.

One key idea of the scattering transform is to generate `first-order' statistics, in contrast to higher-order moments, which multiply an increasing number of field intensities and cause instability to outliers. Being a linear operator, the wavelet convolution certainly keeps the `first-order' property. However, for a homogeneous random field, convolution alone cannot extract information beyond the mean of the original field $\langle I\rangle$, because the expected value operator commutes with all linear transformations. Extracting more information requires non-linear operations. For example, in $N$-point functions, the multiplication of field intensities plays the role. The scattering transform, on the other hand, employs the \textit{modulus} operation, which is a natural choice to preserve the desired property of working with first-order statistics \citep{Mallat_2012}.

As the modulus is taken in the real space and is non-linear, its behaviour in Fourier space is not simple. Nevertheless, we collect some intuitive understandings and present them in Appendix \ref{app:ST_Fourier} for interested readers.

\subsection{Information extraction beyond the power spectrum}
\label{sec:beyond_P}

There are a number of similarities between the power spectrum and each single iteration of the scattering transform. Indeed, the power spectrum can be defined using the formalism of the 1st-order scattering coefficients $S_1=\langle|I_0\star\psi|\rangle$:
\begin{eqnarray}
    P(\bm{k}) &\propto& \langle |I_0 \star \psi'|^2 \rangle\; {\rm with}\;\psi'=\text{e}^{-i\bm{k}\cdot\bm{x}}\;.
\label{eq:P}
\end{eqnarray}
The differences between the two estimators $S_1$ and $P(k)$ are the choice of convolution kernels (wavelets $\psi$ or Fourier modes $\psi'$) and that of the norm (L$^1$ versus L$^2$). Therefore, the 1st-order scattering coefficients have similarity to the power spectrum. Both of them characterize the strength of fluctuations (or clustering) as a function of scale.

However, in the case of the power spectrum, the convolution kernel ($\psi'$~=~$e^{-i\bm{k}\cdot\bm{x}}$) is completely de-localized in real space. Thus, the power spectrum's version of $I_1$ fields ($|I_0 \star \psi'|^2$) lose all spatial information. In contrast, the use of localized wavelets in the scattering transform allows $I_1$ to preserve spatial information, as shown in Figure~\ref{fig:I} and \ref{fig:cos_diff}. 
According to the analogy with the power spectrum, the mean of an $I_1$ field characterizes the average amplitude of Fourier modes selected by the wavelets, whereas the spatial distribution of fluctuations in $I_1$, missing in the power spectrum analogue, in turn encodes the phase interaction between those Fourier modes. This information can be extracted by applying the scattering operations once again, $I_1\rightarrow I_2= |I_1\star\psi_2|=\left| |I_0\star\psi_1|\star\psi_2\right|$, and then measuring the mean of $I_2$, i.e., 2nd-order scattering coefficients $S_2$.

According to the power spectrum analogy, $S_2$ coefficients resemble the power spectrum of $I_1$ fields and measure clustering properties on $I_1$. Because $I_1$ fields highlight the regions where fluctuations around a scale are stronger, the 2nd-order coefficients can be understood as measuring the clustering of structures highlighted in $I_1$, i.e., the `clustering of (clustered) structures'. 

This leads to an interesting intuition: we need two points to describe the scale of one structure and an additional two points for another one. Therefore, the 2nd-order scattering coefficients $S_2$, which measure the clustering of sized structures, include information up to about 4-point. In general, an $n$th-order scattering coefficient $S_n$ will contain information up to about $2^n$-point function of the input field. By this `hierarchical clustering' design, the scattering-transform expansion quickly includes information from higher-order statistics.

However, it should be noted that there is still a fundamental difference between the scattering transform and $N$-point functions. There are mainly two difficulties associated with $N$-point functions to characterize a random field: the failure to describe distribution tails and the huge number of configurations. The first difficulty, related to the multiplication of multiple random variables, leads to high variances and also prevents the extraction of information from fields whose pdf has a tail \citep{Carron_2011}. The scattering transform, which uses modulus and does not enhance the tail, can significantly alleviate this problem. We will discuss it further in another paper (Cheng et al. in prep.). The second difficulty may be alleviated by an efficient binning. For example, the hierarchical wavelet transform used in the scattering transform \emph{is} a binning strategy that can also be applied to $N$-point functions (see Appendix \ref{app:ST_Fourier}).

\section{Application in weak lensing cosmology}
\label{sec:WL_application}

We now show that the scattering transform can be a powerful tool in observational cosmology to extract non-Gaussian information from the matter density field. To illustrate this point, we consider an application with 2-dimensional fields: we show how well cosmological parameters can be constrained using the scattering coefficients of weak lensing convergence maps $\kappa(\vec\theta)$ or, equivalently, measurements of cosmic shear. Being projections of the density field along the line-of-sight, these maps present an appreciable level of non-gaussianities on scales smaller than a few degrees, reflecting the non-linear growth of matter fluctuations. For the necessary background on cosmology with gravitational lensing, we refer the reader to reviews \citep{Kilbinger_2015, Mandelbaum_2018}. 

We explore the use of our reduced scattering coefficients on simulated weak lensing convergence maps to infer $\Omega_\text{m}$ and $\sigma_8$ and compare their performance with that of the power spectrum. We also compare our results with that of a state-of-the-art CNNs by \citet{Ribli_2019} and peak count statistics.

\subsection{Simulated convergence maps}
\label{sec:data}

We use mock convergence maps generated by the Columbia Lensing team\footnote{\href{https://http://columbialensing.org}{http://columbialensing.org}} and described in \citet{Matilla_2016} and \citet{Gupta_2018}. The maps are produced through ray-tracing to redshift $z$ = 1 in the output of dark-matter-only $N$-body simulations for a set of $\Lambda$-CDM cosmologies. Each simulation is run in a 240~$h^{-1}$~Mpc box with 512$^3$ particles. The cosmologies differ only in two parameters: the present matter density relative to the critical density $\Omega_\text{m}$, and a normalization of the power spectrum $\sigma_8$. Other cosmological parameters are fixed: baryon density $\Omega_b$~=~0.046, Hubble constant $h$~=~0.72, scalar spectral index $n_s$~=~0.96, effective number of relativistic degrees of freedom $n_\text{eff}$~=~3.04, and neutrino masses $m_\nu$~=~0.0. The dark energy density is set so that the universe is spatially flat, i.e., $\Omega_\Lambda$ = $1-\Omega_\text{m}$. For each cosmology, 512 convergence maps with 3.5$\times$3.5 deg$^2$ field of view are generated from the simulations, allowing us to sample cosmic variance. The corresponding scales are well suited to probing the non-Gaussianities of the convergence field \citep{Kilbinger_2015}. These maps were also used by \citet{Ribli_2019}. To compare our results to \citet{Ribli_2019}, we use the same resolution as theirs, down-sampling the original 1024$^2$ pixel maps to a 512$^2$ resolution with 0.41 arcmin per pixel.

\subsection{Galaxy shape noise and smoothing}
\label{sec:shape_noise}

In practice, convergence or shear estimates are obtained from measurements of galaxy shapes, with a level of noise that depends on the galaxy ellipticity distribution and their number density on the sky. To first order, background galaxies used for shear measurements have a wide range of redshifts and are not correlated. The noise can be well approximated as Gaussian white noise. Its contribution to the convergence maps can be modelled \citep{vanWaebeke_2000} as
\begin{align}
    \sigma^2_\text{noise} = \frac{\sigma^2_\epsilon}{2 n_g A_\text{pix}}\,,
\end{align}
where $\sigma_\epsilon^2$ is the intrinsic variance of ellipticity of galaxies, which is taken to be 0.4$^2$, $n_g$ is the number density of background galaxies, $A_\text{pix}$ is the area per pixel, which is 0.1682 arcmin$^2$. For some existing and on-going surveys such as CFHTLenS, KiDS\footnote{\href{ http://kids.strw.leidenuniv.nl}{\url{ http://kids.strw.leidenuniv.nl}}}, and DES\footnote{\href{https://www.darkenergysurvey.org}{\url{https://www.darkenergysurvey.org}}}, $n_g$ is around 10 arcmin$^{-2}$ \citep{Kilbinger_2013, DESDR1_2018}; for some upcoming surveys we expect substantially higher densities: $n_g\sim $ 25 arcmin$^{-2}$ for the survey at Vera C. Rubin Observatory (LSST) \footnote{\href{https://www.lsst.org}{\url{https://www.lsst.org}}}, $n_g>$ 30 arcmin$^{-2}$ for \textit{Euclid}\footnote{\href{https://sci.esa.int/euclid}{\url{https://sci.esa.int/euclid}}}, and $n_g\sim$ 50--75 arcmin$^{-2}$ for the planned survey with Nancy Grace Roman Space Telescope (\textit{WFIRST})\footnote{\href{https://roman.gsfc.nasa.gov}{\url{https://roman.gsfc.nasa.gov}}}. 

After adding noise, we also smooth the maps. As the power of Gaussian white noise is distributed mostly at high frequencies, smoothing the convergence maps can help to increase the signal-to-noise of specific estimators. By default, we do not smooth the noiseless maps, and we perform a $\sigma$ = 1 arcmin (2.44 pixel) Gaussian smoothing on noisy maps.

\subsection{Statistical descriptors}
\label{sec:descriptors}

\textit{Scattering coefficients}: For each 3.5$\times$3.5 deg$^2$ convergence field in each cosmology, we apply the scattering transform up to 2nd order using the `kymatio' python package\footnote{\href{https://www.kymat.io}{https://www.kymat.io}} \citep{andreux2020kymatio} and then calculate the reduced coefficients ($s_0$, $s_1$, $s_2$) as defined in Section~\ref{sec:st_formulation}. To probe the available range of scales, we set $J$ = 8 and $L$ = 4 in the scattering transform, i.e., we use 8 scales spaced logarithmically with central wavelengths between 1.2 arcmin and 75 arcmin and 4 azimuthal orientations, resulting in 32 different wavelets used in total. 

\begin{figure*}
    \centering
    \includegraphics[width=0.3\textwidth]{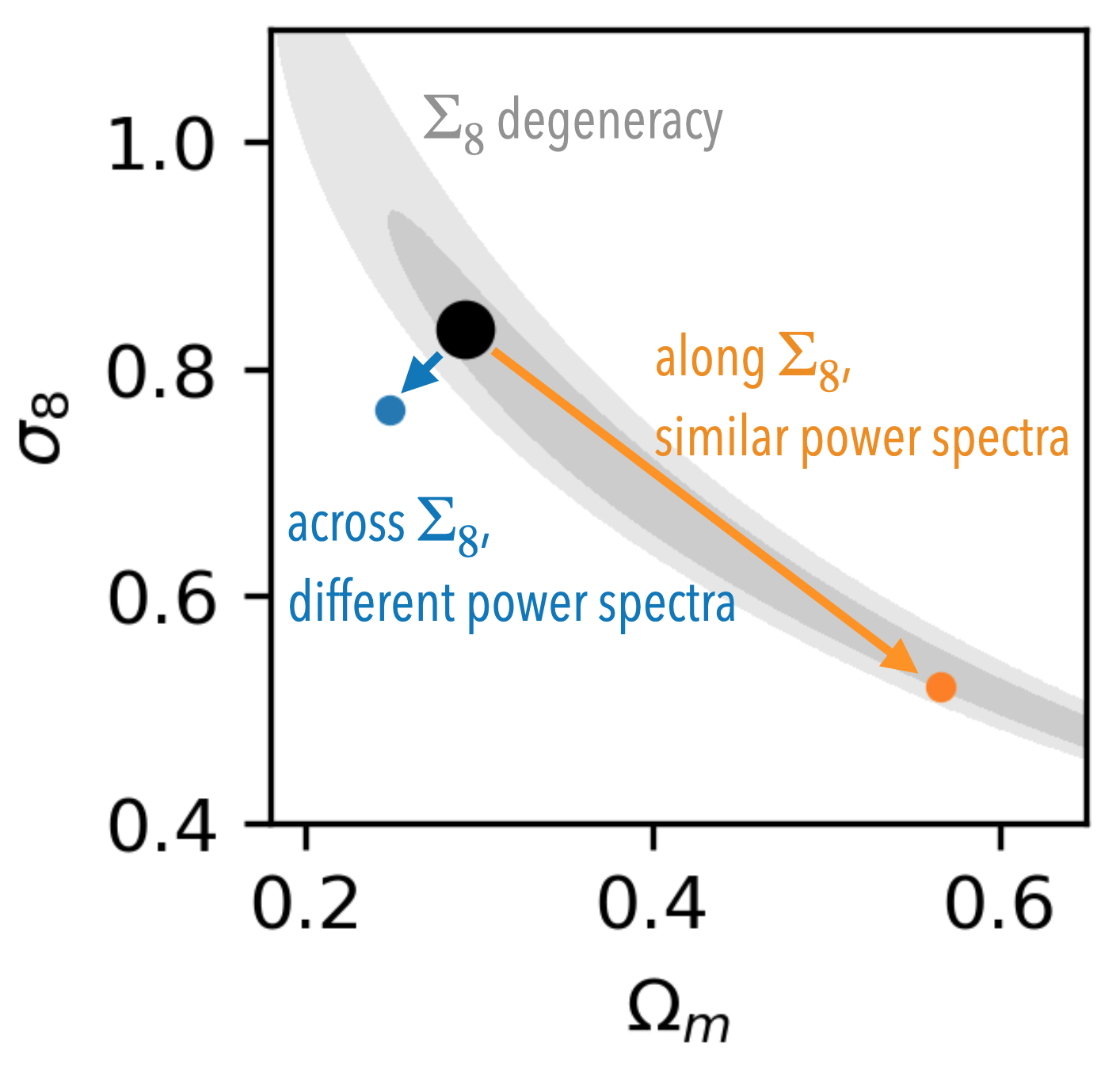}~~~~~~~~~~~~~~~~~~~~~~~~~~~
    \includegraphics[width=0.4\textwidth]{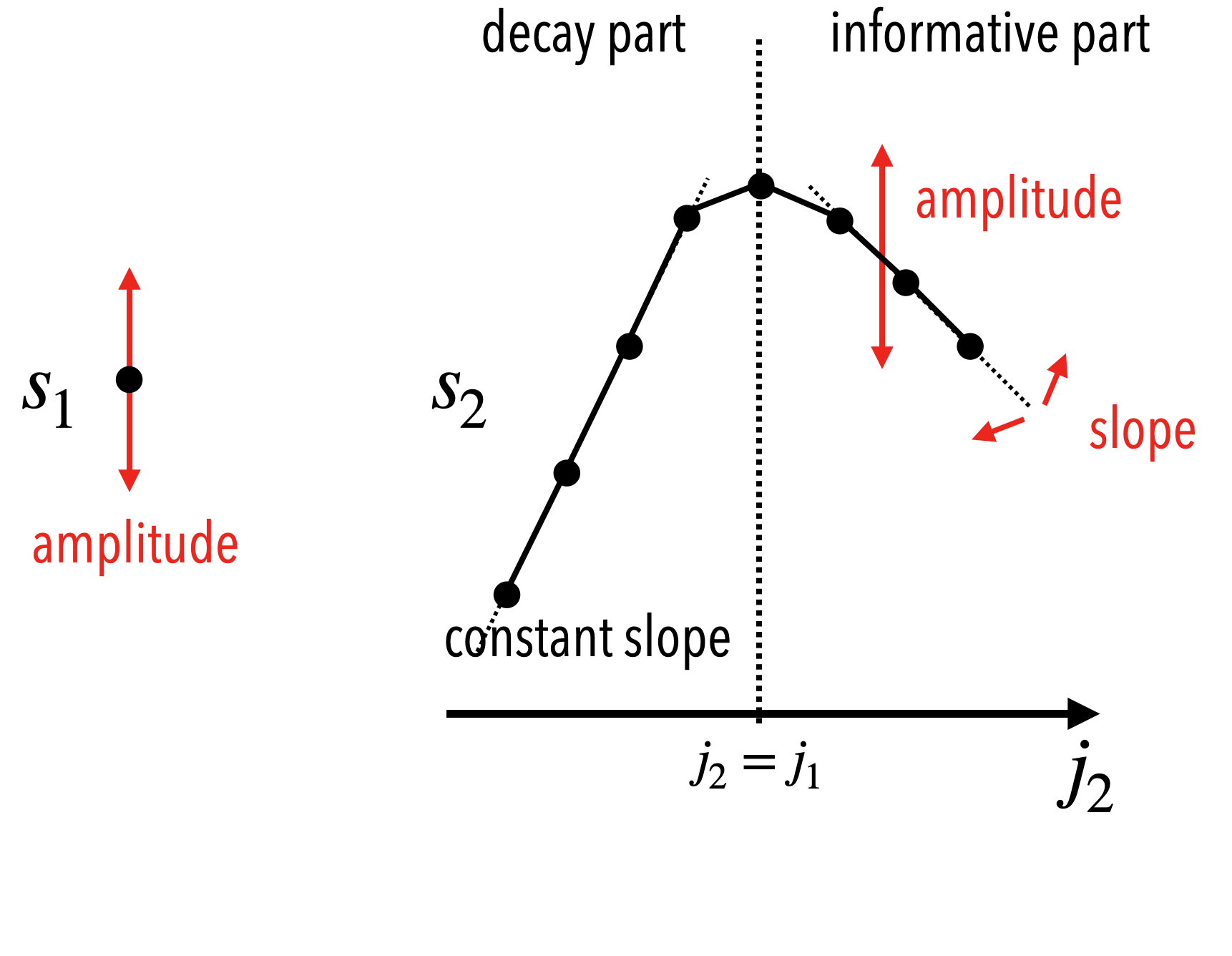}
    \includegraphics[width=\textwidth]{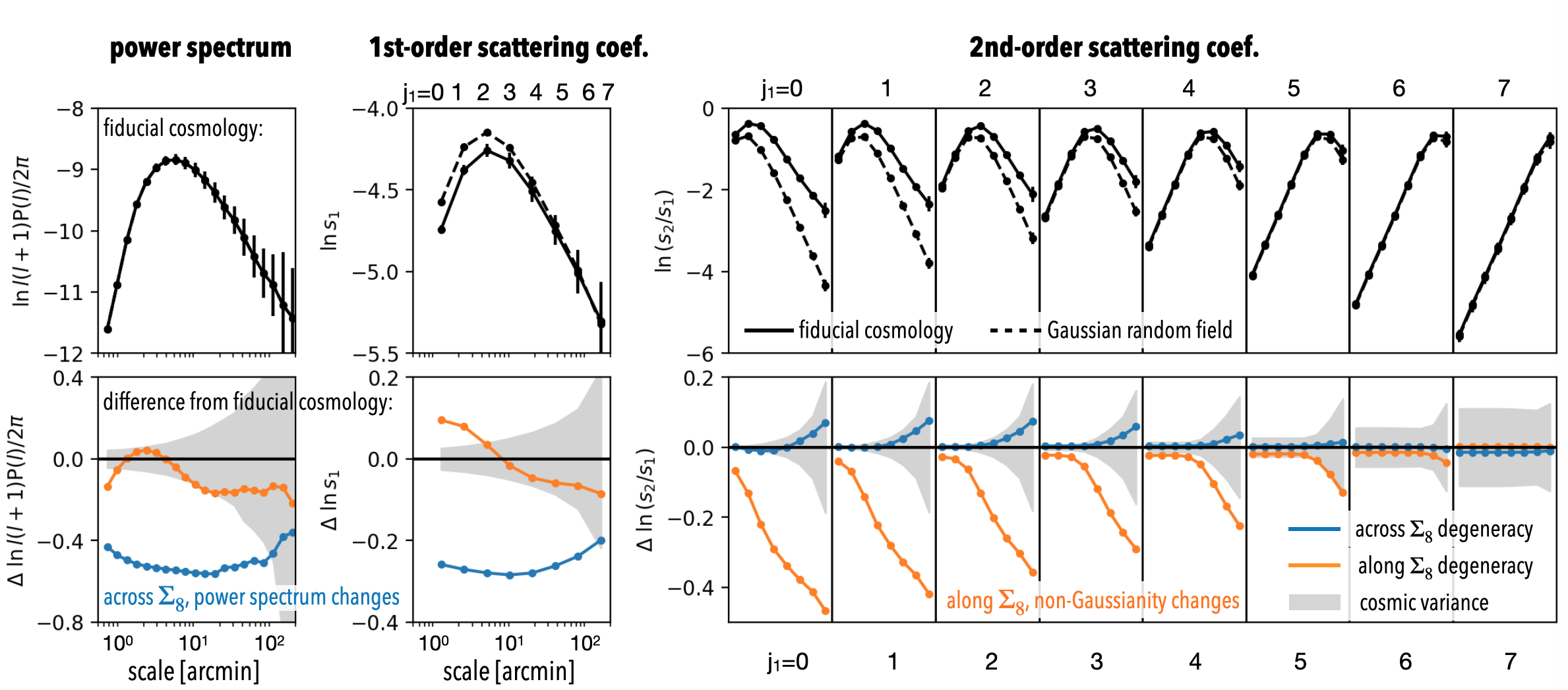}
    \caption{
    \textit{Upper-left Panel}: The fiducial cosmology (black) and two other cosmologies on the ($\Omega_\text{m},\sigma_8$) plane. \textit{Upper-right Panel}: Illustration of reduced scattering coefficients $s_1(j_1)$ and $s_2(j_1,j_2)$ for a single $j_1$ scale. \textit{Lower Panel}: The power spectrum and scattering coefficients for the three cosmologies in noiseless case. The first row presents coefficients of the fiducial cosmology and of Gaussian random fields with the same power spectrum, and the second row shows changes of coefficients ($\Delta$ coef.) when we move from the fiducial cosmology to the other two. Error bars and gray shaded regions show cosmic variance, i.e., the variability among realizations. The 1st-order scattering coefficients behave similarly to the power spectrum, while the 2nd-order scattering coefficients can break the $\Sigma_8$ degeneracy, along which non-Gaussianity of weak lensing field changes.}
    \label{fig:cosmological_dependence}
\end{figure*}

By default, the `kymatio' package only calculate the 2nd-order coefficients with $j_2>j_1$, because the coefficients with $j_2\leq j_1$ is mainly determined by the property of wavelets but not the input field, as illustrated by the upper-right sketch of Figure~\ref{fig:cosmological_dependence}. Intuitively, this is because structures of a particular size, say $j_1$, do not have meaningful clustering at scales smaller than their own size. A mathematical reasoning for this property can also be found in Appendix \ref{app:ST_Fourier}.
To demonstrate these coefficients' behaviour, we modified the `kymatio' code to calculate them, and show them together with the coefficients with $j_2>j_1$ in Figure~\ref{fig:cosmological_dependence}. Nevertheless, we checked that they do not contribute to constraining cosmological parameters, and therefore in our inference analysis, we only use 2nd-order coefficients with $j_2>j_1$, which yields an even more compact set of 1 + 8 + 28 = 37 scattering coefficients used for our cosmological inference.

\textit{Power spectrum}: For the same set of input fields, we also compute the power spectrum and peak count statistics using the publicly available `LensTools' python package\footnote{\href{https://lenstools.readthedocs.io}{https://lenstools.readthedocs.io}} \citep{Petri_2016}. The power spectrum is calculated within 20 bins in the range 100~$\leq l\leq$~37500 (corresponding to 0.58--216 arcmin) with logarithmic spacing, following the setting adopted in \citet{Ribli_2019}. 

\textit{Peak count}: In our analysis, a peak is defined as a pixel with higher convergence ($\kappa$) than its eight neighbors. Then, peaks are binned by their $\kappa$ values and counted in each bin. We adopt a binning similar to that in \citet{LPH15}. We use 20 bins in total, including 18 bins linearly spaced between $\kappa$ = --0.02 and 0.12, one bin for peaks below --0.02, and one bin above 0.12. For reference, $\kappa$ = 0.12 corresponds to a significance of peak $\nu\equiv\kappa/\sigma_\text{noise}$ around 7 when $n_g$ = 30. Although using more bins for very high peaks ($\kappa>$0.12) may enhance the constraining power of the peak count method, we do not use them in this study, because the count distribution of these rare peaks can no longer be approximated by Gaussian distribution \citep[e.g.,][]{Lin_2015}.

To obtain constraints on the cosmological parameters, we use the Fisher inference framework \citep{Fisher_1935, Tegmark_1997}, in which we assume the probability distribution of statistical descriptors is a multivariate Gaussian distribution for a given cosmology. The mean vector and covariance matrix of this Gaussian distribution are dependent on cosmological parameters and estimated from the 512 realizations of each cosmology in simulations. Details of our cosmological inference framework are described in Appendix \ref{app:inference}. Because $s_1$, $s_2$, and power spectra must be positive for a non-trivial field, we consider their logarithm to better satisfy a multivariate Gaussian likelihood. To perform the cosmological inference analysis with the three methods introduced above, we use
\begin{itemize}
    \item 37 scattering coefficients
    \item 20 power spectrum coefficients
    \item 20 peak count coefficients.
\end{itemize}

\section{Results}
\label{sec:results}

In this section, we examine the distribution and cosmological sensitivity of scattering coefficients, and present their constraining power for two cosmological parameters, $\Omega_\text{m}$ and $\sigma_8$. We show that the scattering coefficients provide substantially more information than the power spectrum and is on a par with CNN. 

\subsection{Cosmological sensitivity of the scattering coefficients}
\label{sec:sensitivity}

In Figure~\ref{fig:cosmological_dependence}, we present the distributions of reduce scattering transform in the noiseless case together with the power spectrum. In the first row, we show the values for a fiducial cosmology that has the Planck cosmology of $\Omega_\text{m}$ = 0.309 and $\sigma_8$ = 0.816 \citep{Planck_2016}. The expected values of these descriptors are estimated by averaging over different realizations of a given cosmology. Error bars, which are the sample standard deviations of realizations, represent the cosmic variance in this noiseless case. We can see the similarity between the power spectrum and $s_1$ coefficients, as they have similar physical meanings (Section~\ref{sec:beyond_P}). We can also see the different behaviours of $s_2$ coefficients for $j_2<j_1$ and $j_2>j_1$, as discussed in Section~\ref{sec:descriptors}.

Then, we investigate the cosmological sensitivity of the power spectrum and scattering coefficients. The power spectrum is known to be mostly sensitive to one combination of the cosmological parameters, namely 
\begin{equation}
    \Sigma_8\equiv \sigma_8 \left(\frac{\Omega_\text{m}}{0.3}\right)^a\;,
\label{eq:Sigma8}
\end{equation}
with $a$ around 0.6 \citep[e.g.,][]{Kilbinger_2015}, but can hardly distinguish cosmologies with the same $\Sigma_8$, as illustrated in the upper panel of Figure~\ref{fig:cosmological_dependence}. Breaking this degeneracy requires the extraction of non-Gaussian information from lensing maps.

In the second row of Figure~\ref{fig:cosmological_dependence}, we show the response of coefficients as cosmological parameters move along (orange curves) and across (blue curves) the $\Sigma_8$ degeneracy. Gray areas indicate cosmic variance of the fiducial cosmology. As expected, the 1st-order scattering coefficients show a cosmological sensitivity similar to that of the power spectrum, because both of them measure the strength of fluctuations as a function of scale.

The 2nd-order scattering coefficients, on the other hand, characterize the spatial distribution of sized fluctuations. To make the 2nd-order scattering coefficients less correlated with the 1st-order ones, here we present de-correlated 2nd-order coefficients $s_2/s_1$, as each $s_2(j_1,j_2)$ is proportional to the corresponding $s_1(j_1)$ according to their definitions \citep{Bruna_2015}. 
These $s_2/s_1$ exhibit particularly high sensitivity to cosmological change along the $\Sigma_8$ degeneracy. In addition, they are indifferent to the other direction of cosmological change, which means they provide a piece of information roughly orthogonal to that carried by the 1st-order coefficients $s_1$ or the power spectrum. 
In noisy cases, though the information from $s_2/s_1$ is not orthogonal to $s_1$ anymore, we have checked that they still provide substantial sensitivity along the $\Sigma_8$ degeneracy. 
Due to this additional sensitivity, the scattering transform can be used to better constrain cosmological parameters than the power spectrum.

\begin{figure}
    \centering
    \includegraphics[width=\columnwidth]{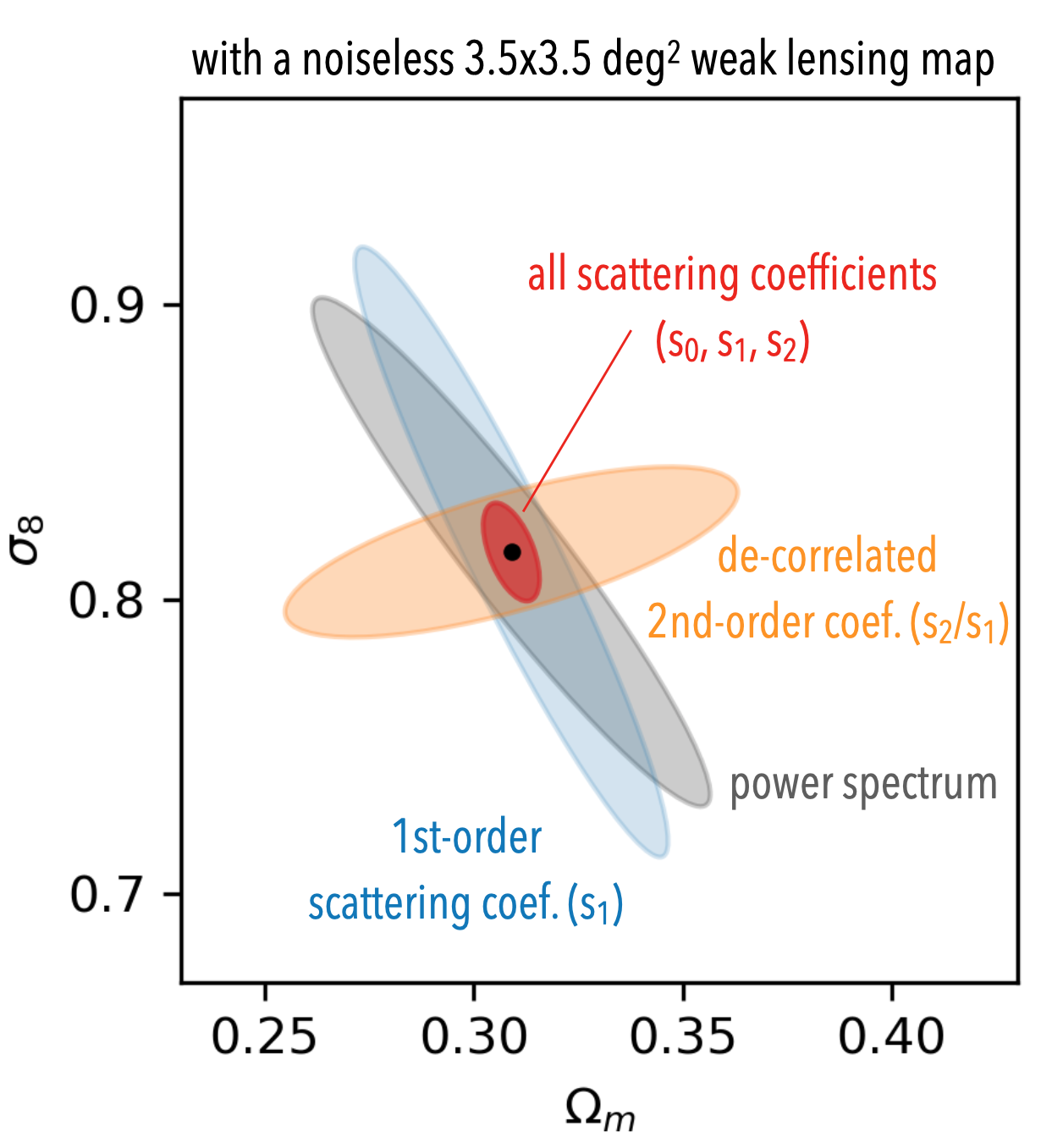}
    \caption{The 1$\sigma$ Fisher forecast of cosmological parameters from a 3.5$\times$3.5 deg$^2$ noiseless convergence map with 0.41 arcmin per pixel resolution. 
    The de-correlated 2nd-order scattering coefficients $s_2/s_1$ provide critical information to break the $\Sigma_8$ degeneracy along which the power spectrum cannot distinguish, therefore drastically improve the constraint.
    }
    \label{fig:noiseless}
\end{figure}

\begin{figure*}
    \centering
    \includegraphics[width=0.9\textwidth]{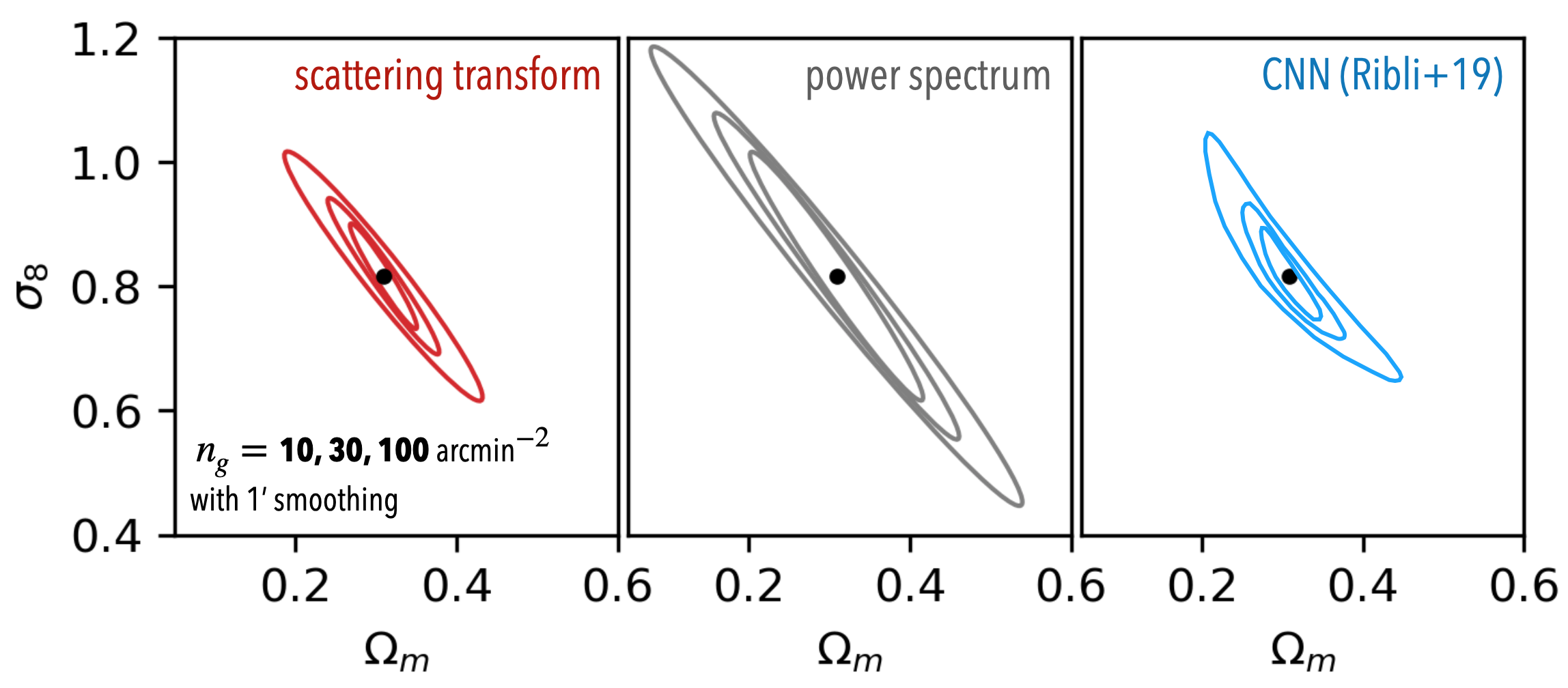}
    \caption{The 1$\sigma$ Fisher forecast of cosmological parameters ($\Omega_\text{m}$ and $\sigma_8$ ) from different descriptors of a 3.5$\times$3.5 deg$^2$ convergence map smoothed with $\sigma$ = 1' Gaussian filter. The scattering coefficients have comparable performance as a state-of-the-art CNN \citep{Ribli_2019} at all noise levels, and 3--5 times better than the power spectrum depending on the noise level.
    }
    \label{fig:comparison}
\end{figure*}

\begin{table*}
 \caption{\label{tab:comparison}Comparison of the constraining power for ($\Omega_\text{m}$, $\sigma_8$) between different methods, with a single 3.5$\times$3.5 deg$^2$ convergence map. The figure of merit is defined as the reciprocal of the 1$\sigma$ confident area based on Fisher matrix (or the 68\% posterior contour, in parentheses) on the ($\Omega_\text{m}$,$\sigma_8$) plane. The convergence maps are smoothed with $\sigma$ = 1' Gaussian filter except for the case shown in the last column with no smoothing.}
 \begin{tabular}{c|ccccc}
  \hline
    Methods & & & $\Omega_\text{m}$--$\sigma_8$ Figure of Merit & \\
    & $n_g$ = 10 arcmin$^{-2}$& $n_g$ = 30 arcmin$^{-2}$& $n_g$ = 100 arcmin$^{-2}$ & noiseless & noiseless (no smoothing)\\
  \hline
    scattering transform: $s_0$ + $s_1$ + $s_2$ & 50 & 140 & 329 & 1053 & 3367\\
    scattering transform: $s_0$ + $s_1$ & 21 & 55 & 133 & 492 & 565\\
    scattering transform: $s_1$ + $s_2$ & 39 & 91 & 181 & 446 & 1720\\
    &&&&&\\
    power spectrum ${\text P}(l)$ & 20 & 40 & 67 & 104 & 253\\
    peak count & 30 & 89 & 162 & 170 & 667\\
    CNN \citep{Ribli_2019} & (44) & (121) & (292) & (1201) & ( - )\\
  \hline
 \end{tabular}
\end{table*}

\subsection{Constraining cosmological parameters}
\label{sec:constraints}

We now present the cosmological constraints set by the scattering coefficients measured from a single 3.5$\times$3.5 deg$^2$ field. For reference, we note that LSST will generate about 2,000 times more data, leading to constraints about 40 times tighter than the numbers presented below. In this study, we only probe the constraints on $\Omega_\text{m}$ and $\sigma_8$ and leave the work of using scattering coefficients to constrain the dark energy equation of state parameter $w$ or neutrino mass $M_\nu$ to future study. Cosmological inference is just another aspect of the cosmological sensitivity problem examined in the previous subsection. The Fisher inference formalism we use in this study is described in Appendix \ref{app:inference}.

We first present results in the noiseless case. In Figure~\ref{fig:noiseless} we demonstrate the 1$\sigma$ Fisher forecast of $\Omega_\text{m}$ and $\sigma_8$ using all scattering coefficients (red ellipse) and power spectrum (gray ellipse). The scattering coefficients provide a dramatically tighter constraint than the power spectrum. We also show a break-down of this constraining power into contributions from 1st-order (blue ellipse) and 2nd-order (orange ellipse) coefficients alone. As expected, the 1st-order coefficients ($s_1$) and power spectrum set similar constraints. The slight difference of ellipse orientation originates from the difference between the ${\rm L}^1$ and ${\rm L}^2$ norms used by the scattering transform and the power spectrum. The de-correlated 2nd-order scattering coefficients ($s_2/s_1$) provide a strong constraint along the $\Sigma_8$ degeneracy, consistent with our cosmological sensitivity discussion in Section~\ref{sec:sensitivity}.

The 0th-order coefficient $s_0$ is the mean of the 3.5$\times$3.5 deg$^2$ field. While its expectation value over the sky is zero, it does carry relevant information on those scales by capturing larger-scale modulations of the convergence field.
We also note that it has strong correlations with other scattering coefficients (and the power spectrum), which is a sign of being in the non-linear regime of cosmology \citep[e.g.,][]{Li_2020}. Therefore, although the expected value of $s_0$ is identically zero in all cosmology, combining $s_0$ with other coefficients helps to substantially tighten the constraints on cosmological parameters. However, this piece of information may not scale as fast with the increasing field of view as the small-scale information, because in real data each patch of 3.5$\times$3.5 deg$^2$ fields on the sky are not independent. The mass sheet degeneracy \citep[e.g.,][]{Bradac_2004} is another problem for using $s_0$, though the $s_0$ of small patches may be obtained by inheriting the zero-point solution of the whole survey. We find that including $s_0$ only improves the constraint of $\Sigma_8$, consistent with the understanding that it is a leakage of larger-scale fluctuation. Similar improvement is also found when combining $s_0$ with the power spectrum.

To be more quantitative, we compare different methods using the reciprocal of the area of their 1$\sigma$ Fisher forecast ellipses on the ($\Omega_\text{m},\sigma_8$) plane as the figure of merit (FoM). In the noiseless case, combining all scattering coefficients ($s_0$, $s_1$, $s_2$) leads to a constraint that is 14 times tighter than that of the power spectrum, 5 times tighter than peak count statistics, and 3.3 times tighter than the joint constraint from power spectrum and peak count.

We then compare the performance of the scattering transform to a state-of-the-art CNN analysis by \citet{Ribli_2019}. To perform a meaningful comparison, we follow \citet{Ribli_2019} to use noiseless convergence maps smoothed with a $\sigma$ = 1 arcmin Gaussian filter. Interestingly, we find that the scattering coefficients extract a similar amount of cosmological information to the CNN trained in \citet{Ribli_2019}. The corresponding figures of merit are shown in Table~\ref{tab:comparison}. \footnote{We note that \citet{Ribli_2019} do include the field mean information in their CNN training. So, a fair comparison would be $s_1+s_2$ versus power spectrum, and $s_0+s_1+s_2$ versus CNN.}

We now consider convergence fields in the presence of galaxy shape noise. As the noise level increases, small-scale structures, which carry plenty of cosmological information, get erased. As a result, the constraining power of the scattering coefficients (as well as other methods) degrades. In Figure~\ref{fig:comparison} we show the Fisher forecast of $\Omega_\text{m}$ and $\sigma_8$ from a 3.5$\times$3.5 deg$^2$ convergence map under three noise levels, using the scattering coefficients and the power spectrum. We also show the posterior constraints from CNNs trained by \citep{Ribli_2019} on the same simulations. The figures of merit for these methods, together with the peak count method, are listed in Table~\ref{tab:comparison}. Again, we find that the scattering transform not only outperforms the power spectrum and peak count, but also provides cosmological constraints on a par with state-of-the-art CNNs.

\begin{figure}
    \centering
    \includegraphics[width=\columnwidth]{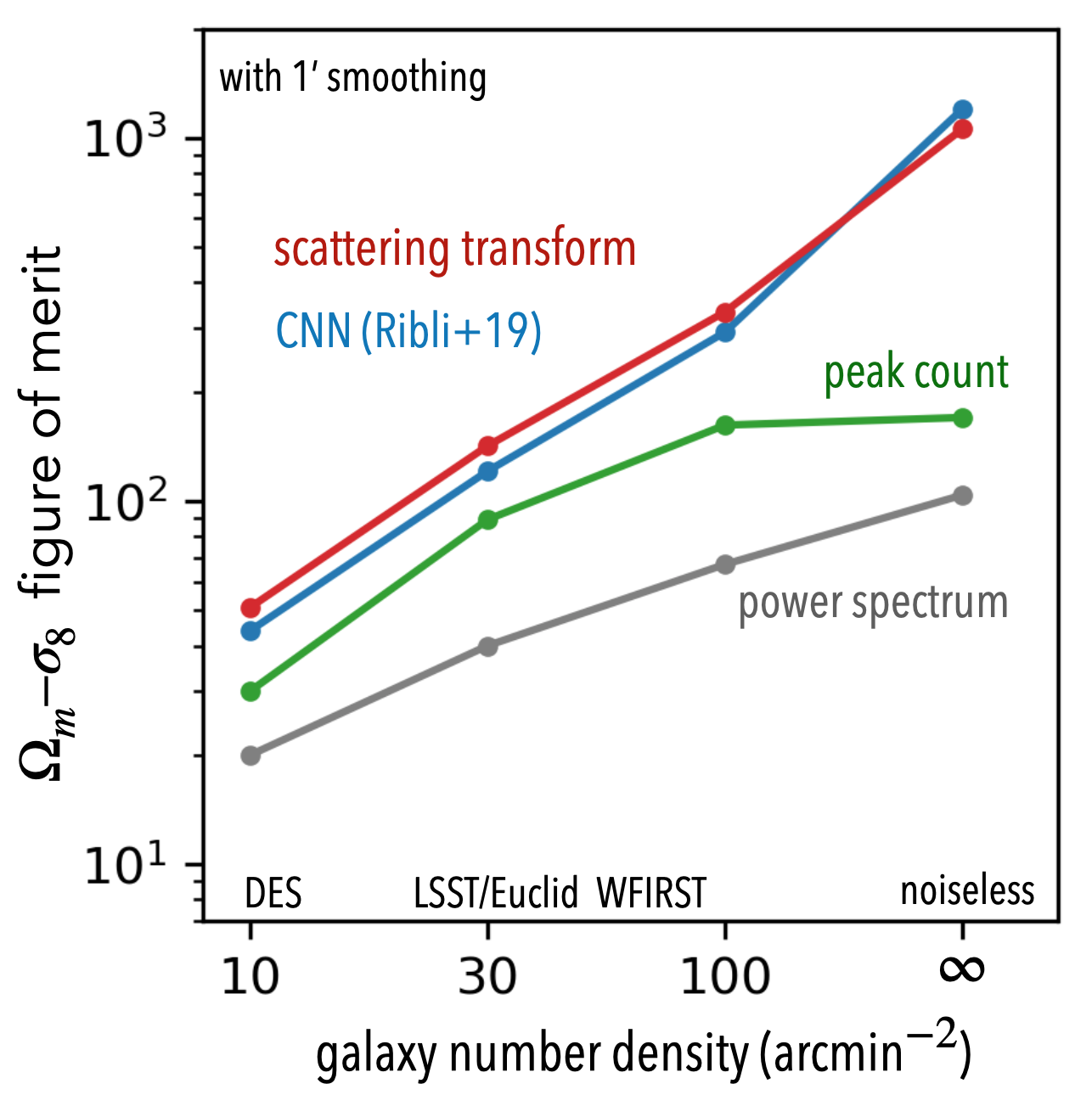}
    \caption{Dependence of the ($\Omega_\text{m},\sigma_8$) constraints with different methods on galaxy shape noise. The figure of merit (FoM) is defined as the 1$\sigma$ confident area on the ($\Omega_\text{m},\sigma_8$) plane. Note that the CNN result \citep{Ribli_2019} is reported in terms of posterior, while others are Fisher forecast. For noisy cases the difference between scattering transform and CNN is not intrinsic but due to the difference between posterior and Fisher forecast.
    Peak count's performance does not increase as fast because it is more sensitive to smoothing scale than the other methods.}
    \label{fig:noise_scaling}
\end{figure}

To summarise, we have demonstrated the power of the scattering transform for cosmological parameter inference with weak lensing data. For simplicity, we focused on the convergence field but a similar analysis can also be performed on the shear field. In Figure~\ref{fig:noise_scaling}, we present quantitative comparisons between the four techniques discussed in our study. It shows the high performance of the scattering transform over a wide range of noise levels. We therefore advocate using this new estimator in the analysis of existing and upcoming weak lensing surveys, in observational cosmology, and more generally, in the analysis of stochastic fields encountered in physics.

\section{Discussion}
\label{sec:discussion}

\subsection{Inference for non-Gaussian fields}


In physics, many inference problems concern estimating physical parameters from realizations of random fields. Ideally, one would like to use the likelihood function of the field itself, but this is often out of reach except for several simple cases such as some Gaussian random fields. Therefore, for the inference problem to be feasible, a statistical representation of the data is often used. Statistical descriptors reduce the dimensionality of the data vector and they tend to Gaussianize according to the central limit theorem. Both of these properties help to regularize the likelihood. However, it is still challenging to find a proper representation because in general a random field can be random in too many different ways. In these cases, a useful characterization must be one that makes use of known properties of the field.


Viewed in this direction, traditional statistical approaches with their own representation framework may or may not suit the properties of particular fields. For example, the peak count statistic used in weak lensing cosmology suits the presence of distinct haloes in convergence maps. $N$-point functions, closely related to perturbation theory and convenient for analytical prediction, represent the field with a series expansion, which makes them good descriptors for fields slightly deviating from a Gaussian one. 
A highly non-Gaussian field, however, requires using larger $N$. As the number of coefficients and the complexity of configurations increase rapidly with $N$, 
$N$-point functions quickly become an inefficient and non-robust representation of the input field.
On the other hand, CNNs try to learn the field properties and search for informative representation through a training optimization. 

Fortunately, the non-Gaussian fields that originate from physical interactions do often have common properties.
Such fields typically display localized, coherent structures in multi-scales, and smaller structures often act as building blocks of larger structures. These properties can be used as the `domain knowledge' to guide our design or choice of the statistical representation in a general sense.
As we will explain in the next section, the design and operations of the scattering transform leads to an efficient and robust representation for such fields, because they are tailored for these properties.

\subsection{Attractive properties of the scattering transform}
\label{sec:attractive}

\textbf{Efficiency}: All the three elements (wavelet convolution, modulus, and the hierarchical design) play essential roles to make the scattering transform efficient.
The use of wavelets balances the resolution in real and frequency domain. As a result, the scattering transform can capture localized information from a large range of scales with only a few coefficients, at each order. 
After selecting structures of scale $j_n$ in one order, the scattering transform then selects structures `assembled' by these $j_n$-scale structures in the next order. This hierarchical design allows the $n$th order scattering coefficients to quickly access configurations described by about $2^n$ points. Moreover, the `low-order' non-linear operator, modulus, helps to collect information even beyond the access of $2^n$ point functions. We will discuss it further in another paper (Cheng et al. in prep.).


These strategies concentrate relevant information to a limited set of statistical descriptors, which is desirable in terms of compactness of the representation and the signal-to-noise ratio of each estimator. For example, in our case, the scattering transform compresses weak lensing information into 37 coefficients, a number that is much smaller than typical bi-spectrum descriptors, while achieving CNN-like constraint on cosmological parameters. 

\textbf{Robustness}: All scattering coefficients are `first-order' statistics in the sense that they are proportional to the input field, and it is proved that the scattering representation is non-expansive, i.e. the distance between two vectors in the scattering representation never exceeds their distance in the original pixel-based representation \citep{Mallat_2012, Bruna_2013}. Therefore, it does not amplify the process variability. This is in contrast to the $N$-point correlation function approach, which requires multiplying an increasing number of field fluctuations and causes high variability. As a result, the scattering coefficients are low variance descriptors and insensitive to outliers. 

The locality of wavelets, which is related to their logarithmic spacing and widths in frequency space, also introduces stability to deformations \citep{Mallat_2012}, which is a desired property of robust descriptors that classical $N$-point functions do not have.

\textbf{Interpretability}: As discussed in Section~\ref{sec:beyond_P}, the scattering coefficients have a simple and intuitive interpretation. They describe clustering properties of the field in the following way:

The 1st-order scattering coefficients are similar to a coarsely binned power spectrum, which characterize the clustering strength at different scales $j_1$. As the scattering transform uses an L$^1$ norm as opposed to an L$^2$ norm, the ratio between $s_1$ coefficients and the power spectrum provides a measure of sparsity of the field. This explains why in Figure~\ref{fig:noiseless} the constraints from 1st-order coefficients and the power spectrum are slightly different, and just combining these two can also provide a stronger constraint on cosmology than using power spectrum alone.

The 2nd-order scattering coefficients characterize the clustering strength of $j_1$-scale structures separated by $j_2$-scales. In other words, these coefficients characterize the clustering of structures selected over a given frequency range, or the `clustering of clustering'. Their departure from their Gaussian counterparts is a robust measure of the strength of non-Gaussianities. The $n$th-order scattering coefficients, though not shown explicitly in this study, can in turn be understood as the strengths of $n$th-order hierarchy of clustering of the field at all different combinations of scales.

\subsection{Comparison to CNNs}

The scattering transform and CNNs share a number of properties. Both of them have hierarchical layers with localized convolution kernels and use a simple non-expansive non-linear operation. Although CNNs are usually trained to directly map a field to physical parameters, their inside can be considered as composed of a convolutional part that extracts spatial features and a second part that learns the mapping from these features to physical parameters. Both parts are trainable and trained together. The scattering transform, on the other hand, uses preset wavelets as convolutional kernels and just a few layers (in our case two layers). So it can be viewed as a non-trainable mini-CNN playing the role of the first part of trainable CNNs. In the scattering transform's approach, the second part of trainable CNNs is supplanted by using traditional regression techniques.

The trainable kernels make CNNs more flexible and may lead to a higher performance for finer classification problems such as classifying different types of rabbits, but in the mean time this over-parametrization defines a much more brittle statistical model \citep{Szegedy_2013, bruna2019multiscale}. Our results imply that compared to CNNs, the scattering transform has enough expressiveness to characterize the matter density field in the cosmological context while holding provable stability properties.
Indeed, as shown by \citet{Ribli_2019}, a CNN trained on convergence maps internally generates kernels similar to (azimuthally averaged) Morlet wavelets. Our results also imply that much of the power of CNNs may be detached from its trainable nature.

Over-parametrized models tend to over-fit, i.e., to `remember' single realizations instead of comprehending the overall property of the whole training set. Thus the over-parametrized CNNs require a large number of simulations as training set to alleviate the over-fitting problem. In contrast, the scattering transform uses preset kernels, thus has no parametrization in the kernels. In addition, the choice of CNN architecture can modify the results substantially, as can be seen in the comparison between results of \citet{Ribli_2019} and \citet{Gupta_2018}. As such, CNNs usually require much, and often ad-hoc, fine-tuning. The scattering transform, on the other hand, is not subject to these sources of variability. It requires the use of simulations only to probe the cosmic variance of the descriptors. 
Without learning the kernels, the scattering transform also significantly save calculation time.

Another view on the over-fitting problem is given by the framework of maximum-entropy regularized estimation, which looks for the most `non-committal' statistical model under the constraints of a `feature vector' of sufficient statistics \citep{jaynes1957information}. There is thus a tension in the design of such vector of sufficient statistics \citep{bruna2019multiscale}: On the one hand, the features should be descriptive enough so that they introduce enough constraints, i.e., typical samples from the estimated model should also be typical in the true distribution; On the other hand, one would like the features to be efficiently estimated from the available samples, so that the corresponding statistical model is robust under resampling. In other words, typical samples from the true distribution should remain typical under the estimated statistical model.

Finally, when applied to observational data, the scattering transform holds another advantage over CNNs, namely the possibility to investigate systematic effects. As traditional statistics, the scattering coefficients can be used to derive not only the best-fitting cosmological parameters, but also an evaluation of the goodness of fit and therefore a sanity check of the result.
In contrast, although the internal machinery of CNNs can be roughly divided into a feature extraction part and a regression one, the CNNs are trained as a whole to learn a direct mapping from the data to the physical parameters. Due to the over-parametrization nature, outputs from intermediate layers (i.e. the intermediate abstraction of CNN) do not typically have good statistical properties. Therefore, when using CNN, it is challenging to check for systematic error in real data.


\subsection{Relation to peak count method}

The non-linear gravitational evolution of density fluctuations in the universe gives rise to haloes, which are virialized systems locally bound by gravity. As highlighted by \citet{Ribli_2019} in their Figure~10, a substantial amount of non-Gaussian cosmological information can be extracted from these features. The peak count method directly captures information in the abundance of haloes. However, it does not characterize the spatial information, including profiles and positions of these haloes, which is also sensitive to cosmological parameters. The scattering transform implicitly extracts a comprehensive information of the abundance, profile, and distribution of haloes by first highlighting structures of particular scales and then characterizing their clustering at other scales, as described in Section~\ref{sec:beyond_P}.
In the limit of small $j_1$ and large $j_2$, the 2nd-order scattering coefficients can be understood as a measure of the `two-halo term' in the halo model at scale $j_2$, weighted by the halo response to the first wavelet with scale $j_1$. This response is related to halo profiles. In general, the scattering transform provides a non-parametric description of the one-halo, two-halo, and transitional regime where haloes overlap and form larger haloes.

\section{Conclusion}
\label{sec:conclusion}

Characterizing arbitrary non-Gaussian fields is challenging as the dimensionality of their description can be arbitrarily high. The subset of fields relevant in physics, however, tends to be more constrained as they typically display localized, coherent structures. In the cosmological context, the matter density field presents another characteristic property, namely hierarchical clustering. An efficient statistical descriptor of the cosmological density field would ideally make use of
these properties.

In this paper, we advocate the use of the scattering transform \citep{Mallat_2012, Bruna_2013}, which generates statistics designed to extract information from complex fields with provable stability properties. It involves operations similar to those found in convolutional neural networks (CNNs): it uses wavelet convolution, which is particularly suitable for characterizing localized structures; it uses modulus as the non-linear operation; and it iterates these operations. However, in contrast to CNNs, the scattering transform does \emph{not} require training. It generates a compact set of robust coefficients, which forms a representation of the input field and can be used as efficient summary statistics for non-Gaussian information. 

We applied the scattering transform to a parameter inference problem in the context of weak lensing cosmology. For simplicity, we focused on the convergence field but a similar analysis can also be performed on the shear field. We used simulated convergence maps generated by ray-tracing $N$-body simulation results \citep{Matilla_2016, Gupta_2018} and measured their scattering coefficients to infer the cosmological parameters $\Omega_\text{m}$ and $\sigma_8$. On maps with and without galaxy shape noise, the scattering transform outperforms the power spectrum and peak counts, and is on par with state-of-the-art CNNs.

As described in section~\ref{sec:attractive}, the scattering transform possesses a series of attractive properties for parameter estimation. It is efficient, robust, and interpretable. Obtained by iteratively applying wavelet convolution and modulus and finally taking the expectation value, the scattering coefficients can be interpreted as the strength of a hierarchy of clustering at various combinations of scales.
Different from $N$-point functions, all scattering coefficients have the welcome property that they remain proportional to the input field, thus avoid instability problems
and extract much more information when the field distribution has a long tail.
Similar to classic statistical estimators, the scattering transform requires no training or tuning and offers the possibility to investigate systematic errors potentially present with real data.

In this paper we demonstrated applications of the scattering transform in weak lensing data. Using it with existing and upcoming surveys (e.g. DES, LSST, \textit{Euclid}, \textit{WFIRST}) can be of great interest to improve constraints and provide consistency checks. Based on its properties and design, the scattering transform can also be an attractive approach for many other applications: in observational cosmology, astrophysics, and beyond.

\section*{Acknowledgements}
We thank the anonymous referee, Jean-Fran\c{c}ois Cardoso, Yi-Kuan Chiang, and Zuhui Fan for useful comments. We also thank Dezs\H{o} Ribli for discussions. 
We thank the Columbia Lensing group
(\href{http://columbialensing.org}{\url{http://columbialensing.org}})
for making their suite of simulated maps available, and NSF for supporting the creation of those maps through grant AST-1210877 and XSEDE allocation AST-140041. YST is supported by the NASA Hubble Fellowship grant HST-HF2-51425.001 awarded by the Space Telescope Science Institute. This work is partially supported by the Alfred P. Sloan Foundation, NSF RI-1816753, NSF CAREER CIF 1845360, NSF CHS-1901091, Samsung Electronics, and the Institute for Advanced Study. SC thanks Siyu Yao for her constant encouragement and inspiration.

\section*{Data availability}

The data underlying this article were accessed from the Columbia Lensing group (\href{http://columbialensing.org}{\url{http://columbialensing.org}}). The derived data generated in this research will be shared on reasonable request to the corresponding author.

\bibliographystyle{mnras}
\bibliography{ST_mnras}

\begin{thebibliography}{}
\makeatletter
\relax
\def\mn@urlcharsother{\let\do\@makeother \do\$\do\&\do\#\do\^\do\_\do\%\do\~}
\def\mn@doi{\begingroup\mn@urlcharsother \@ifnextchar [ {\mn@doi@}
  {\mn@doi@[]}}
\def\mn@doi@[#1]#2{\def\@tempa{#1}\ifx\@tempa\@empty \href
  {http://dx.doi.org/#2} {doi:#2}\else \href {http://dx.doi.org/#2} {#1}\fi
  \endgroup}
\def\mn@eprint#1#2{\mn@eprint@#1:#2::\@nil}
\def\mn@eprint@arXiv#1{\href {http://arxiv.org/abs/#1} {{\tt arXiv:#1}}}
\def\mn@eprint@dblp#1{\href {http://dblp.uni-trier.de/rec/bibtex/#1.xml}
  {dblp:#1}}
\def\mn@eprint@#1:#2:#3:#4\@nil{\def\@tempa {#1}\def\@tempb {#2}\def\@tempc
  {#3}\ifx \@tempc \@empty \let \@tempc \@tempb \let \@tempb \@tempa \fi \ifx
  \@tempb \@empty \def\@tempb {arXiv}\fi \@ifundefined
  {mn@eprint@\@tempb}{\@tempb:\@tempc}{\expandafter \expandafter \csname
  mn@eprint@\@tempb\endcsname \expandafter{\@tempc}}}

\bibitem[\protect\citeauthoryear{{Abbott} et~al.,}{{Abbott}
  et~al.}{2018}]{DESDR1_2018}
{Abbott} T.~M.~C.,  et~al., 2018, \mn@doi [\apjs] {10.3847/1538-4365/aae9f0},
  \href {https://ui.adsabs.harvard.edu/abs/2018ApJS..239...18A} {239, 18}

\bibitem[\protect\citeauthoryear{{Allys}, {Levrier}, {Zhang}, {Colling},
  {Regaldo-Saint Blancard}, {Boulanger}, {Hennebelle}  \& {Mallat}}{{Allys}
  et~al.}{2019}]{Allys_2019}
{Allys} E.,  {Levrier} F.,  {Zhang} S.,  {Colling} C.,  {Regaldo-Saint
  Blancard} B.,  {Boulanger} F.,  {Hennebelle} P.,   {Mallat} S.,  2019,
  \mn@doi [\aap] {10.1051/0004-6361/201834975}, \href
  {https://ui.adsabs.harvard.edu/abs/2019A&A...629A.115A} {629, A115}

\bibitem[\protect\citeauthoryear{{Allys}, {Marchand}, {Cardoso},
  {Villaescusa-Navarro}, {Ho}  \& {Mallat}}{{Allys} et~al.}{2020}]{Allys_2020}
{Allys} E.,  {Marchand} T.,  {Cardoso} J.~F.,  {Villaescusa-Navarro} F.,  {Ho}
  S.,   {Mallat} S.,  2020, arXiv e-prints, \href
  {https://ui.adsabs.harvard.edu/abs/2020arXiv200606298A} {p. arXiv:2006.06298}

\bibitem[\protect\citeauthoryear{{And\'en} \& {Mallat}}{{And\'en} \&
  {Mallat}}{2011}]{AndenMallat_2011}
{And\'en} J.,  {Mallat} S.,  2011, International Society for Music Information
  Retrieval Conference, \href
  {https://www.di.ens.fr/data/publications/papers/ismir-final.pdf} {pp
  657--662}

\bibitem[\protect\citeauthoryear{{And\'en} \& {Mallat}}{{And\'en} \&
  {Mallat}}{2014}]{AndenMallat_2014}
{And\'en} J.,  {Mallat} S.,  2014, \mn@doi [IEEE Transactions on Signal
  Processing] {10.1109/TSP.2014.2326991}, \href
  {https://arxiv.org/abs/1304.6763} {62, 4114}

\bibitem[\protect\citeauthoryear{Andreux et~al.,}{Andreux
  et~al.}{2020}]{andreux2020kymatio}
Andreux M.,  et~al., 2020, Journal of Machine Learning Research, \href
  {https://ui.adsabs.harvard.edu/abs/2018arXiv181211214A} {21, 1}

\bibitem[\protect\citeauthoryear{{Bernardeau}, {Mellier}  \& {van
  Waerbeke}}{{Bernardeau} et~al.}{2002}]{Bernardeau2002}
{Bernardeau} F.,  {Mellier} Y.,   {van Waerbeke} L.,  2002, \mn@doi [\aap]
  {10.1051/0004-6361:20020700}, \href
  {https://ui.adsabs.harvard.edu/abs/2002A&A...389L..28B} {389, L28}

\bibitem[\protect\citeauthoryear{{Brada{\v{c}}}, {Lombardi}  \&
  {Schneider}}{{Brada{\v{c}}} et~al.}{2004}]{Bradac_2004}
{Brada{\v{c}}} M.,  {Lombardi} M.,   {Schneider} P.,  2004, \mn@doi [\aap]
  {10.1051/0004-6361:20035744}, \href
  {https://ui.adsabs.harvard.edu/abs/2004A&A...424...13B} {424, 13}

\bibitem[\protect\citeauthoryear{{Bruna} \& {Mallat}}{{Bruna} \&
  {Mallat}}{2013}]{Bruna_2013}
{Bruna} J.,  {Mallat} S.,  2013, \mn@doi [IEEE Transactions on Pattern Analysis
  and Machine Intelligence] {doi: 10.1109/TPAMI.2012.230}, \href
  {https://arxiv.org/abs/1203.1513} {35, 1872}

\bibitem[\protect\citeauthoryear{Bruna \& Mallat}{Bruna \&
  Mallat}{2019}]{bruna2019multiscale}
Bruna J.,  Mallat S.,  2019, Mathematical Statistics and Learning, \href
  {https://arxiv.org/abs/1801.02013} {1, 257}

\bibitem[\protect\citeauthoryear{{Bruna}, {Mallat}, {Bacry}  \& J.-F.}{{Bruna}
  et~al.}{2015}]{Bruna_2015}
{Bruna} J.,  {Mallat} S.,  {Bacry} E.,   J.-F. M.,  2015, \mn@doi [The Annals
  of Statistics] {10.1214/14-AOS1276}, \href {https://arxiv.org/abs/1311.4104}
  {43, 323}

\bibitem[\protect\citeauthoryear{{Carron}}{{Carron}}{2011}]{Carron_2011}
{Carron} J.,  2011, \mn@doi [\apj] {10.1088/0004-637X/738/1/86}, \href
  {https://ui.adsabs.harvard.edu/abs/2011ApJ...738...86C} {738, 86}

\bibitem[\protect\citeauthoryear{{Carron} \& {Szapudi}}{{Carron} \&
  {Szapudi}}{2013}]{Carron_2013}
{Carron} J.,  {Szapudi} I.,  2013, \mn@doi [\mnras] {10.1093/mnras/stt1215},
  \href {https://ui.adsabs.harvard.edu/abs/2013MNRAS.434.2961C} {434, 2961}

\bibitem[\protect\citeauthoryear{{Eickenberg}, {Exarchakis}, {Hirn}, {Mallat}
  \& {Thiry}}{{Eickenberg} et~al.}{2018}]{Eickenberg_2018}
{Eickenberg} M.,  {Exarchakis} G.,  {Hirn} M.,  {Mallat} S.,   {Thiry} L.,
  2018, \mn@doi [\jcp] {10.1063/1.5023798}, \href
  {https://ui.adsabs.harvard.edu/abs/2018JChPh.148x1732E} {148, 241732}

\bibitem[\protect\citeauthoryear{Fisher}{Fisher}{1935}]{Fisher_1935}
Fisher R.~A.,  1935, \mn@doi [Journal of the Royal Statistical Society]
  {10.2307/2342435}, 98, 39

\bibitem[\protect\citeauthoryear{{Fu} et~al.,}{{Fu} et~al.}{2014}]{Fu_2014}
{Fu} L.,  et~al., 2014, \mn@doi [\mnras] {10.1093/mnras/stu754}, \href
  {https://ui.adsabs.harvard.edu/abs/2014MNRAS.441.2725F} {441, 2725}

\bibitem[\protect\citeauthoryear{{Gama}, {Ribeiro}  \& {Bruna}}{{Gama}
  et~al.}{2018}]{gama2018diffusion}
{Gama} F.,  {Ribeiro} A.,   {Bruna} J.,  2018, arXiv e-prints, \href
  {https://ui.adsabs.harvard.edu/abs/2018arXiv180608829G} {p. arXiv:1806.08829}

\bibitem[\protect\citeauthoryear{{Giblin} et~al.,}{{Giblin}
  et~al.}{2018}]{Giblin_2018}
{Giblin} B.,  et~al., 2018, \mn@doi [\mnras] {10.1093/mnras/sty2271}, \href
  {https://ui.adsabs.harvard.edu/abs/2018MNRAS.480.5529G} {480, 5529}

\bibitem[\protect\citeauthoryear{{Gupta}, {Matilla}, {Hsu}  \&
  {Haiman}}{{Gupta} et~al.}{2018}]{Gupta_2018}
{Gupta} A.,  {Matilla} J. M.~Z.,  {Hsu} D.,   {Haiman} Z.,  2018, \mn@doi
  [\prd] {10.1103/PhysRevD.97.103515}, \href
  {https://ui.adsabs.harvard.edu/abs/2018PhRvD..97j3515G} {97, 103515}

\bibitem[\protect\citeauthoryear{{Hartlap}, {Simon}  \& {Schneider}}{{Hartlap}
  et~al.}{2007}]{Hartlap_2007}
{Hartlap} J.,  {Simon} P.,   {Schneider} P.,  2007, \mn@doi [\aap]
  {10.1051/0004-6361:20066170}, \href
  {https://ui.adsabs.harvard.edu/abs/2007A&A...464..399H} {464, 399}

\bibitem[\protect\citeauthoryear{{Hikage} et~al.,}{{Hikage}
  et~al.}{2003}]{Hikage_2003}
{Hikage} C.,  et~al., 2003, \mn@doi [\pasj] {10.1093/pasj/55.5.911}, \href
  {https://ui.adsabs.harvard.edu/abs/2003PASJ...55..911H} {55, 911}

\bibitem[\protect\citeauthoryear{{Hirn}, {Mallat}  \& {Poilvert}}{{Hirn}
  et~al.}{2017}]{Hirn_2017}
{Hirn} M.,  {Mallat} S.,   {Poilvert} N.,  2017, \mn@doi [Multiscale Modeling
  \& Simulation] {10.1137/16M1075454}, \href
  {https://ui.adsabs.harvard.edu/abs/2016arXiv160504654H} {15, 827–863}

\bibitem[\protect\citeauthoryear{{Jain} \& {Van Waerbeke}}{{Jain} \& {Van
  Waerbeke}}{2000}]{Jain_2000}
{Jain} B.,  {Van Waerbeke} L.,  2000, \mn@doi [\apjl] {10.1086/312480}, \href
  {https://ui.adsabs.harvard.edu/abs/2000ApJ...530L...1J} {530, L1}

\bibitem[\protect\citeauthoryear{{Jaynes}}{{Jaynes}}{1957}]{jaynes1957information}
{Jaynes} E.~T.,  1957, \mn@doi [Physical Review] {10.1103/PhysRev.106.620},
  \href {https://ui.adsabs.harvard.edu/abs/1957PhRv..106..620J} {106, 620}

\bibitem[\protect\citeauthoryear{{Kilbinger}}{{Kilbinger}}{2015}]{Kilbinger_2015}
{Kilbinger} M.,  2015, \mn@doi [Reports on Progress in Physics]
  {10.1088/0034-4885/78/8/086901}, \href
  {https://ui.adsabs.harvard.edu/abs/2015RPPh...78h6901K} {78, 086901}

\bibitem[\protect\citeauthoryear{{Kilbinger} et~al.,}{{Kilbinger}
  et~al.}{2013}]{Kilbinger_2013}
{Kilbinger} M.,  et~al., 2013, \mn@doi [\mnras] {10.1093/mnras/stt041}, \href
  {https://ui.adsabs.harvard.edu/abs/2013MNRAS.430.2200K} {430, 2200}

\bibitem[\protect\citeauthoryear{{Kratochvil}, {Haiman}  \& {May}}{{Kratochvil}
  et~al.}{2010}]{Kratochvil_2010}
{Kratochvil} J.~M.,  {Haiman} Z.,   {May} M.,  2010, \mn@doi [\prd]
  {10.1103/PhysRevD.81.043519}, \href
  {https://ui.adsabs.harvard.edu/abs/2010PhRvD..81d3519K} {81, 043519}

\bibitem[\protect\citeauthoryear{{Kratochvil}, {Lim}, {Wang}, {Haiman}, {May}
  \& {Huffenberger}}{{Kratochvil} et~al.}{2012}]{Kratochvil_2012}
{Kratochvil} J.~M.,  {Lim} E.~A.,  {Wang} S.,  {Haiman} Z.,  {May} M.,
  {Huffenberger} K.,  2012, \mn@doi [\prd] {10.1103/PhysRevD.85.103513}, \href
  {https://ui.adsabs.harvard.edu/abs/2012PhRvD..85j3513K} {85, 103513}

\bibitem[\protect\citeauthoryear{{Lecun}, {Bottou}, {Bengio}  \&
  {Haffner}}{{Lecun} et~al.}{1998}]{lecun-98}
{Lecun} Y.,  {Bottou} L.,  {Bengio} Y.,   {Haffner} P.,  1998, \mn@doi
  [Proceedings of the IEEE] {10.1109/5.726791}, 86, 2278

\bibitem[\protect\citeauthoryear{{Li}, {Dodelson}  \& {Croft}}{{Li}
  et~al.}{2020}]{Li_2020}
{Li} P.,  {Dodelson} S.,   {Croft} R. A.~C.,  2020, \mn@doi [\prd]
  {10.1103/PhysRevD.101.083510}, \href
  {https://ui.adsabs.harvard.edu/abs/2020PhRvD.101h3510L} {101, 083510}

\bibitem[\protect\citeauthoryear{{Lin} \& {Kilbinger}}{{Lin} \&
  {Kilbinger}}{2015}]{Lin_2015}
{Lin} C.-A.,  {Kilbinger} M.,  2015, \mn@doi [\aap]
  {10.1051/0004-6361/201526659}, \href
  {https://ui.adsabs.harvard.edu/abs/2015A&A...583A..70L} {583, A70}

\bibitem[\protect\citeauthoryear{{Liu}, {Petri}, {Haiman}, {Hui}, {Kratochvil}
  \& {May}}{{Liu} et~al.}{2015a}]{LPH15}
{Liu} J.,  {Petri} A.,  {Haiman} Z.,  {Hui} L.,  {Kratochvil} J.~M.,   {May}
  M.,  2015a, \mn@doi [\prd] {10.1103/PhysRevD.91.063507}, \href
  {https://ui.adsabs.harvard.edu/abs/2015PhRvD..91f3507L} {91, 063507}

\bibitem[\protect\citeauthoryear{{Liu} et~al.,}{{Liu} et~al.}{2015b}]{LPL15}
{Liu} X.,  et~al., 2015b, \mn@doi [\mnras] {10.1093/mnras/stv784}, \href
  {https://ui.adsabs.harvard.edu/abs/2015MNRAS.450.2888L} {450, 2888}

\bibitem[\protect\citeauthoryear{Mallat}{Mallat}{2010}]{Mallat10recursiveinterferometric}
Mallat S.,  2010, in Proceedings of 18th European Signal Processing Conference,
  Denmark. pp 716--720

\bibitem[\protect\citeauthoryear{{Mallat}}{{Mallat}}{2012}]{Mallat_2012}
{Mallat} S.,  2012, \mn@doi [Communications on Pure and Applied Mathematics]
  {10.1002/cpa.21413}, \href {https://arxiv.org/abs/1101.2286} {65, 1331}

\bibitem[\protect\citeauthoryear{{Mandelbaum}}{{Mandelbaum}}{2018}]{Mandelbaum_2018}
{Mandelbaum} R.,  2018, \mn@doi [\araa] {10.1146/annurev-astro-081817-051928},
  \href {https://ui.adsabs.harvard.edu/abs/2018ARA&A..56..393M} {56, 393}

\bibitem[\protect\citeauthoryear{{Marian}, {Smith}  \& {Bernstein}}{{Marian}
  et~al.}{2009}]{Marian_2009}
{Marian} L.,  {Smith} R.~E.,   {Bernstein} G.~M.,  2009, \mn@doi [\apjl]
  {10.1088/0004-637X/698/1/L33}, \href
  {https://ui.adsabs.harvard.edu/abs/2009ApJ...698L..33M} {698, L33}

\bibitem[\protect\citeauthoryear{{Mecke}, {Buchert}  \& {Wagner}}{{Mecke}
  et~al.}{1994}]{Mecke_1994}
{Mecke} K.~R.,  {Buchert} T.,   {Wagner} H.,  1994, \aap, \href
  {https://ui.adsabs.harvard.edu/abs/1994A&A...288..697M} {288, 697}

\bibitem[\protect\citeauthoryear{{Neyrinck}, {Szapudi}  \& {Szalay}}{{Neyrinck}
  et~al.}{2011}]{Neyrinck_2011}
{Neyrinck} M.~C.,  {Szapudi} I.,   {Szalay} A.~S.,  2011, \mn@doi [\apj]
  {10.1088/0004-637X/731/2/116}, \href
  {https://ui.adsabs.harvard.edu/abs/2011ApJ...731..116N} {731, 116}

\bibitem[\protect\citeauthoryear{{Petri}}{{Petri}}{2016}]{Petri_2016}
{Petri} A.,  2016, \mn@doi [Astronomy and Computing]
  {10.1016/j.ascom.2016.06.001}, \href
  {https://ui.adsabs.harvard.edu/abs/2016A&C....17...73P} {17, 73}

\bibitem[\protect\citeauthoryear{{Pisani} et~al.,}{{Pisani}
  et~al.}{2019}]{Pisani_2019}
{Pisani} A.,  et~al., 2019, \baas, \href
  {https://ui.adsabs.harvard.edu/abs/2019BAAS...51c..40P} {51, 40}

\bibitem[\protect\citeauthoryear{{Planck Collaboration} et~al.,}{{Planck
  Collaboration} et~al.}{2016}]{Planck_2016}
{Planck Collaboration} et~al., 2016, \mn@doi [\aap]
  {10.1051/0004-6361/201525830}, \href
  {https://ui.adsabs.harvard.edu/abs/2016A&A...594A..13P} {594, A13}

\bibitem[\protect\citeauthoryear{{Ribli}, {Pataki}  \& {Csabai}}{{Ribli}
  et~al.}{2019a}]{Ribli_2019a}
{Ribli} D.,  {Pataki} B.~{\'A}.,   {Csabai} I.,  2019a, \mn@doi [Nature
  Astronomy] {10.1038/s41550-018-0596-8}, \href
  {https://ui.adsabs.harvard.edu/abs/2019NatAs...3...93R} {3, 93}

\bibitem[\protect\citeauthoryear{{Ribli}, {Pataki}, {Zorrilla Matilla}, {Hsu},
  {Haiman}  \& {Csabai}}{{Ribli} et~al.}{2019b}]{Ribli_2019}
{Ribli} D.,  {Pataki} B.~{\'A}.,  {Zorrilla Matilla} J.~M.,  {Hsu} D.,
  {Haiman} Z.,   {Csabai} I.,  2019b, \mn@doi [\mnras] {10.1093/mnras/stz2610},
  \href {https://ui.adsabs.harvard.edu/abs/2019MNRAS.490.1843R} {490, 1843}

\bibitem[\protect\citeauthoryear{{Sefusatti}, {Crocce}, {Pueblas}  \&
  {Scoccimarro}}{{Sefusatti} et~al.}{2006}]{Sefusatti_2006}
{Sefusatti} E.,  {Crocce} M.,  {Pueblas} S.,   {Scoccimarro} R.,  2006, \mn@doi
  [\prd] {10.1103/PhysRevD.74.023522}, \href
  {https://ui.adsabs.harvard.edu/abs/2006PhRvD..74b3522S} {74, 023522}

\bibitem[\protect\citeauthoryear{{Semboloni}, {Schrabback}, {van Waerbeke},
  {Vafaei}, {Hartlap}  \& {Hilbert}}{{Semboloni} et~al.}{2011}]{Semboloni_2011}
{Semboloni} E.,  {Schrabback} T.,  {van Waerbeke} L.,  {Vafaei} S.,  {Hartlap}
  J.,   {Hilbert} S.,  2011, \mn@doi [\mnras]
  {10.1111/j.1365-2966.2010.17430.x}, \href
  {https://ui.adsabs.harvard.edu/abs/2011MNRAS.410..143S} {410, 143}

\bibitem[\protect\citeauthoryear{{Shirasaki} \& {Yoshida}}{{Shirasaki} \&
  {Yoshida}}{2014}]{Shirasaki_2014}
{Shirasaki} M.,  {Yoshida} N.,  2014, \mn@doi [\apj]
  {10.1088/0004-637X/786/1/43}, \href
  {https://ui.adsabs.harvard.edu/abs/2014ApJ...786...43S} {786, 43}

\bibitem[\protect\citeauthoryear{{Sifre} \& {Mallat}}{{Sifre} \&
  {Mallat}}{2013}]{Sifre_2013}
{Sifre} L.,  {Mallat} S.,  2013, the IEEE Conference on Computer Vision and
  Pattern Recognition (CVPR), \href
  {https://openaccess.thecvf.com/content_cvpr_2013/html/Sifre_Rotation_Scaling_and_2013_CVPR_paper.html}
  {pp 1233--1240}

\bibitem[\protect\citeauthoryear{{Simpson}, {James}, {Heavens}  \&
  {Heymans}}{{Simpson} et~al.}{2011}]{Simpson_2011}
{Simpson} F.,  {James} J.~B.,  {Heavens} A.~F.,   {Heymans} C.,  2011, \mn@doi
  [\prl] {10.1103/PhysRevLett.107.271301}, \href
  {https://ui.adsabs.harvard.edu/abs/2011PhRvL.107A1301S} {107, 271301}

\bibitem[\protect\citeauthoryear{{Sinz}, {Swift}, {Brumwell}, {Liu}, {Kim},
  {Qi}  \& {Hirn}}{{Sinz} et~al.}{2020}]{Sinz_2020}
{Sinz} P.,  {Swift} M.~W.,  {Brumwell} X.,  {Liu} J.,  {Kim} K.~J.,  {Qi} Y.,
  {Hirn} M.,  2020, arXiv e-prints, \href
  {https://ui.adsabs.harvard.edu/abs/2020arXiv200601247S} {p. arXiv:2006.01247}

\bibitem[\protect\citeauthoryear{{Szegedy}, {Zaremba}, {Sutskever}, {Bruna},
  {Erhan}, {Goodfellow}  \& {Fergus}}{{Szegedy} et~al.}{2013}]{Szegedy_2013}
{Szegedy} C.,  {Zaremba} W.,  {Sutskever} I.,  {Bruna} J.,  {Erhan} D.,
  {Goodfellow} I.,   {Fergus} R.,  2013, arXiv e-prints, \href
  {https://ui.adsabs.harvard.edu/abs/2013arXiv1312.6199S} {p. arXiv:1312.6199}

\bibitem[\protect\citeauthoryear{{Takada} \& {Jain}}{{Takada} \&
  {Jain}}{2003}]{Takada_2003}
{Takada} M.,  {Jain} B.,  2003, \mn@doi [\apjl] {10.1086/368066}, \href
  {https://ui.adsabs.harvard.edu/abs/2003ApJ...583L..49T} {583, L49}

\bibitem[\protect\citeauthoryear{{Tegmark}, {Taylor}  \& {Heavens}}{{Tegmark}
  et~al.}{1997}]{Tegmark_1997}
{Tegmark} M.,  {Taylor} A.~N.,   {Heavens} A.~F.,  1997, \mn@doi [\apj]
  {10.1086/303939}, \href
  {https://ui.adsabs.harvard.edu/abs/1997ApJ...480...22T} {480, 22}

\bibitem[\protect\citeauthoryear{Welling}{Welling}{2005}]{Welling_2005}
Welling M.,  2005, AISTATS, \href
  {https://www.semanticscholar.org/paper/Robust-Higher-Order-Statistics-Welling/aa6d3565e8ea47e394877aec1582d6cb8b1ab8eb}
  {p.~7}

\bibitem[\protect\citeauthoryear{{Zorrilla Matilla}, {Haiman}, {Hsu}, {Gupta}
  \& {Petri}}{{Zorrilla Matilla} et~al.}{2016}]{Matilla_2016}
{Zorrilla Matilla} J.~M.,  {Haiman} Z.,  {Hsu} D.,  {Gupta} A.,   {Petri} A.,
  2016, \mn@doi [\prd] {10.1103/PhysRevD.94.083506}, \href
  {https://ui.adsabs.harvard.edu/abs/2016PhRvD..94h3506Z} {94, 083506}

\bibitem[\protect\citeauthoryear{{van Waerbeke}}{{van
  Waerbeke}}{2000}]{vanWaebeke_2000}
{van Waerbeke} L.,  2000, \mn@doi [\mnras] {10.1046/j.1365-8711.2000.03259.x},
  \href {https://ui.adsabs.harvard.edu/abs/2000MNRAS.313..524V} {313, 524}

\makeatother
\end{thebibliography}

\appendix

\section{Morlet Wavelets}
\label{app:Morlet}

Wavelets are localized oscillations in real space and band-pass filters in Fourier space. If we simply use a Gaussian envelope to modulate a plane wave, then we obtain a Gabor function,
\begin{equation}
    G(\bm{x}) = \frac{1}{\sqrt{|\bm{\Sigma}|}}e^{-\bm{x}^T\bm{\Sigma}^{-1}\bm{x}/2}e^{i \bm{k}_0 \cdot \bm{x}}\,,
\end{equation}
where $\bm{\Sigma}$ is the covariance matrix describing the size and shape of the Gaussian envelope, and $\bm{k}_0$ determines the frequency of the modulated oscillation. To keep maximum symmetry, usually $\bm{\Sigma}$ is selected to have only 1 eigen-value different from the others, and $\bm{k}_0$ to be along that eigen-direction. Thus we denote the eigen-value along $\bm{k}_0$ by $\sigma^2$ and the other eigen-value by $\sigma^2/s^2$. The parameter $s$ is also the ratio of transverse to radial width of the wavelet in Fourier space.

The Fourier transform of a Gabor function is simply a Gaussian filter centred at $\bm{k}_0$,
\begin{equation}
    \tilde{G}(\bm{k}) = e^{-(\bm{k}-\bm{k}_0)^T\bm{\Sigma}(\bm{k}-\bm{k}_0)/2}\,.
\end{equation}
Wider envelope in real space makes narrower filter in Fourier space. Note that the product $k_0\sigma$ determines the number of oscillations within $\pm\pi\approx3$ standard deviation of the Gaussian envelope and allows for a trade off between spatial and frequency resolution.

Unfortunately, a Gaussian profile in Fourier space does not go to zero at 0 frequency. This contradicts the admissibility of wavelet which requires wavelets to strictly be band-pass filters, not low-pass filters. Therefore, a small correction is required. A simple solution is to introduce an offset, $\beta$, before the Gaussian modulation. In Fourier space this is equivalent to subtracting another Gaussian profile centred at 0 to cancel out the 0-frequency contribution. Families of wavelets created in this way are called Morlet wavelets. Formally,
\begin{equation}
    \psi(\bm{x}) = \frac{1}{\sqrt{|\bm{\Sigma}|}}e^{-\bm{x}^T\bm{\Sigma}^{-1}\bm{x}/2}\left(e^{i \bm{k}_0 \cdot \bm{x}}-\beta\right)\,,
\end{equation}
where $\beta=e^{-\bm{k}_0^T\bm{\Sigma}\bm{k}_0/2}$ is determined by the admissibility criterion. Its Fourier transform is
\begin{equation}
\tilde{\psi}(\bm{k}) = \tilde{G}(\bm{k}) - \beta e^{-\bm{k}^T\bm{\Sigma}\bm{k}/2}\,.
\end{equation}

In our study, which is a 2-dimension case, we follow the settings used in the `kymatio' package mentioned in Section~\ref{sec:descriptors},
\begin{align}
\label{eq:Morlet_param}
    \sigma&=0.8\times2^j\nonumber\\
    k_0&=\frac{3\pi}{4\times2^j}\nonumber\\
    s&=4/L\,,
\end{align}
where $\sigma$ is in unit of pixels, $j$ is an integer starting from 0, and $k_0$ is always between 0 and 1. This choice allow a family of Morlet wavelets best covers the whole Fourier space with a dyadic sequence of scales ($2^j$). Examples of the Morlet wavelets we use are shown in Figure~\ref{fig:wavelets}. Within the wavelet envelope, there are about 2 cycles of oscillations, because $k_0\sigma\approx2$.

\section{Scattering transform in Fourier space}
\label{app:ST_Fourier}

It is enlightening to collect some intuition of the scattering transform in the Fourier domain. In general, as a non-linear operator, a modulus in real space will mix Fourier modes and scatter information among different frequencies. In particular, taking the modulus of $I\star\psi$, where $\psi$ has a single peak in Fourier space, will re-express $I$'s information around $\psi$'s frequency in forms of lower frequencies. In other words, the typical frequency of $|I\star\psi|$ is lower than $I\star\psi$. 

Intuitively, this is because the modulus is converting complex-valued oscillations into its local strength, namely its envelope. Formally, this can be revealed by first writing $|I\star\psi|$ as $\sqrt{(I\star\psi)(I\star\psi)^* }$, where $^*$ stands for complex conjugate, and then Taylor expanding the square root in terms of $(I\star\psi)(I\star\psi)^* - C$, where $C$ is the mean of $(I\star\psi)(I\star\psi)^*$ over all pixels \citep{Mallat10recursiveinterferometric}. The leading term of the Taylor expansion is proportional to $(I\star\psi)(I\star\psi)^* - C$ itself, which corresponds to $I\star\psi$'s auto-correlation in Fourier space. When the power spectrum of $I$ is a smooth function, the frequency distribution of $I\star\psi$ is  similar to $\psi$.
For the Morlet wavelets used in the scattering transform, the central wavenumber of the wavelet $\psi$ is roughly $k_0$ (as defined in Appendix \ref{app:Morlet}), and its half-width in Fourier space around $1/\sigma$.
So, its auto-correlation will have an half-width around $\sqrt{2}/\sigma$ and a centroid at 0. As $\sqrt{2}/\sigma \approx 0.75 k_0 < k_0$ (Equation~\ref{eq:Morlet_param}), this means that the typical frequency of $|I\star\psi|$ is lower than $I\star\psi$. Therefore, the core operation $I\rightarrow|I\star\psi|$ re-expresses high frequency information of $I_n$ in terms of lower frequency modes including the 0-frequency component in the next-order fields $I_{n+1}$. As the 0-frequency component is translation invariant, it can be directly used as a statistical descriptor of the original field.

Writing the modulus $|x|$ as $\sqrt{|x|^2}=\sqrt{x\cdot x^*}$ brings an interesting question: what happens if we replace each modulus by modulus squared? It can be shown that, in this case, the $n$th-order scattering coefficients will exactly become some averaged 2$^n$-point-spectra weighted (binned) by wavelets. Nevertheless, they are not equivalent to any degenerate case of 2$^n$-point functions in either real or Fourier domain.
For example, at the 2nd order, these `pseudo' scattering coefficients become $\iiint \tilde{I_0}(\bm{k}_1)\tilde{I_0}(-\bm{k}_1'-\bm{k}_2) \tilde{I_0}(\bm{k}_1')\tilde{I_0}(-\bm{k}_1'+\bm{k}_2)
\cdot W \cdot d\bm{k}_1 d\bm{k}_1' d\bm{k}_2$, where the weight is determined by the wavelets: $W = \tilde{\psi_1}(\bm{k}_1) \tilde{\psi_1}(\bm{k}_1+\bm{k}_2)  \tilde{\psi_1}(-\bm{k}_1') \tilde{\psi_1}(-\bm{k}_1'+\bm{k}_2) \tilde{\psi_2}^2(\bm{k}_2)$, and the tilde sign denotes Fourier conjugate. Although these `pseudo' coefficients may help us understand the connection between scattering transform and $N$-point functions in terms of how they organise spatial configurations, the genuine scattering transform is fundamentally different from $N$-point functions, because it generates `first-order' estimators, which alleviates the problem of classic moments described in \citet{Carron_2011} when dealing with tailed probability distribution. Indeed, we find that the constraining power of genuine scattering coefficients is about 4 times stronger than these `pseudo' ones (in the noiseless, unsmoothed case). We will discuss this further in another paper (Cheng et al. in prep.).


\section{Cosmological inference framework}
\label{app:inference}

In this appendix we describe the Fisher forecast formalism used to infer the cosmological parameters in this study. According to the Cram\'er--Rao inequality, the variance of any unbiased estimator $\hat{\bm{\theta}}$ for model parameters $\bm{\theta}$ cannot be smaller than the inverse of the Fisher information matrix $\mathbfss{I}(\bm{\theta})$ of the model:
\begin{align}
    \text{cov}(\hat{\bm{\theta}}) \geq \mathbfss{I}(\bm{\theta})^{-1}.
\end{align}
Elements of the Fisher matrix is defined as
\begin{align}
    \I_{m,n}(\bm{\theta}) \equiv \left\langle \frac{\partial \text{ln}\,p(\bm{x}|\bm{\theta} )}{\partial \theta_m} \frac{\partial \text{ln}\,p(\bm{x}|\bm{\theta} )}{\partial \theta_n}\right\rangle\,,
\end{align}
where $\bm{x}$ is the observable, $p$ is the likelihood function, $\langle\cdot\rangle$ is the expectation over $\bm{x}$. In our cosmological case, $\bm{\theta}$ represents cosmological parameters, $\bm{\theta}=(\Omega_\text{m}, \sigma_8)$, and $\bm{x}$ represents the statistical descriptors such as the scattering coefficients. The function $p(\bm{x}|\bm{\theta})$ is called the likelihood of $\bm{\theta}$ when $\bm{x}$ is fixed, and is called the probability density function (PDF) of $\bm{x}$ when $\bm{\theta}$ is fixed. 

In our study, we assume that given any cosmology $\bm{\theta}$, the PDF of statistical descriptors $\bm{x}$ is Gaussian:
\begin{align}
\label{eq:Gaussian_likelihood}
    p(\bm{x}|\bm{\theta} ) \propto \frac{1}{\sqrt{|\mathbfss{C}|}}\text{exp}[-\frac12 (\bm{x}-\bm{\mu})^T\mathbfss{C}^{-1}(\bm{x}-\bm{\mu})]\,,
\end{align}
where $\mathbfss{C}(\bm{\theta})$ and $\bm{\mu}(\bm{\theta})$ are the mean and covariance matrix depending on the cosmological parameters $\bm{\theta}$. Thus, elements of the Fisher matrix can be written as
\begin{align}
    I_{m,n} =& \frac{\partial \bm{\mu}^T}{\partial \theta_m} \mathbfss{C}^{-1} \frac{\partial \bm{\mu}}{\partial \theta_n} +\nonumber\\ &\frac{1}{2} {\rm tr} (\mathbfss{C}^{-1} \frac{\partial \mathbfss{C}}{\partial \theta_m} \mathbfss{C}^{-1} \frac{\partial \mathbfss{C}}{\partial \theta_n} )\,,\label{eq:Fisher_Gaussian}
\end{align}
where the first and second items describe the information from cosmological dependence of $\bm{\mu}$ and $\mathbfss{C}$, respectively. To obtain these items for arbitrary cosmology, we first calculate the sample mean and covariance matrix of the 512 realizations of each cosmology in the simulations (Section~\ref{sec:data}). The sample mean is an unbiased estimator of the real mean vector, but to unbiasly estimate the inverse of covariance matrix, $\mathbfss{C}^{-1}$, a correction factor is needed \citep{Hartlap_2007}: 
\begin{align}
    \widehat{\mathbfss{C}^{-1}} = \frac{N-D-2}{N-1}\widehat{\mathbfss{C}}^{-1}\,,
\end{align}
where $\widehat{\mathbfss{C}^{-1}}$ is the unbiased estimator in the inverse, $N$ is the number of independent sample used for the estimation, $D$ is the dimension of each data vector, and $\widehat{\mathbfss{C}}$ is the sample covariance before Bessel's correction.
Then, with a further assumption that $\bm{\mu}$ and $\mathbfss{C}$ have smooth cosmological dependence, we use 3rd-order polynomials to fit for the cosmological dependence of $\bm{\mu}$'s elements and use 2nd-order polynomials for $\mathbfss{C}$'s elements.

\bsp	
\label{lastpage}
\end{document}